\documentclass[reprint, aps, prx, superscriptaddress]{revtex4-2}

\usepackage[english]{babel}
\makeatletter
\renewcommand{\selectlanguage}[1]{}

\makeatother

\usepackage{amsmath}
\usepackage{amsfonts}
\usepackage{physics}
\usepackage{graphicx}
\usepackage[breaklinks=true,
            colorlinks=true,
            linkcolor=blue,
            citecolor=blue,
            urlcolor=blue]{hyperref}

\usepackage{etoolbox}
\apptocmd{\sloppy}{\hbadness 10000\relax}{}{}
\usepackage{jabbrv}

\usepackage{enumitem}
\newlist{mylist}{itemize}{1}
\setlist[mylist]{label=\textbf{Step 1:}}

\usepackage{upgreek}

\usepackage{array}
\usepackage{multirow}

\usepackage{contour}
\DeclareMathAlphabet{\mathpzc}{OT1}{pzc}{m}{it}

\usepackage{bm}

\usepackage{comment}

\usepackage[normalem]{ulem}
\usepackage{amsmath}
\usepackage{enumerate}
\usepackage{amsfonts}
\usepackage{epsfig}
\usepackage{mathbbol}

\def\tr{\mathop{\rm tr}}
\def\Im{\mathop{\rm Im} }

\newcommand\half{{\ensuremath{\frac{1}{2}}}}
\newcommand\p{\ensuremath{\partial}}
\newcommand\pp{\ensuremath{\bm\nabla}}

\newcommand{\mE}{\mathpzc{E}}

\newcommand{\mH}{\mathpzc{H}}

\newcommand{\mV}{\mathpzc{V}}

\newcommand{\ff}{{\mathfrak{f}}}

\newcommand{\be}{\begin{equation}}
\newcommand{\ee}{\end{equation}}
\newcommand{\bea}{\begin{eqnarray}}
\newcommand{\eea}{\end{eqnarray}}
\newcommand{\bega}{\begin{gather}}
\newcommand{\eega}{\end{gather}}

\newcommand{\bi}{\begin{itemize}}
\newcommand{\ei}{\end{itemize}}
\newcommand{\ben}{\begin{enumerate}}
\newcommand{\een}{\end{enumerate}}
\newcommand{\bca}{\begin{cases}}
\newcommand{\eca}{\end{cases}}
\newcommand{\bln}{\begin{align}}
\newcommand{\eln}{\end{align}}
\newcommand{\bst}{\begin{split}}
\newcommand{\est}{\end{split}}
\def\ie{\begin{equation}\begin{aligned}}
\def\fe{\end{aligned}\end{equation}}
\newcommand{\bma}{\le(\begin{matrix}}
\newcommand{\ema}{\end{matrix}\ri)}
\newcommand{\bwt}{\begin{widetext}}
\newcommand{\ewt}{\end{widetext}}

\newcommand\lam{\lambda}

\newcommand\Th{{\Theta}}
\def\th{{\theta}}

\newcommand\ov{\over}
\newcommand\ha{{\half}}

\def\le{\left}
\def\ri{\right}

\newcommand\sE{{\ensuremath{{\mathcal E}}}}
\newcommand\sF{{\ensuremath{{\mathcal F}}}}
\newcommand\sI{{\ensuremath{{\mathcal I}}}}

\newcommand\sH{{\ensuremath{{\mathcal H}}}}

\newcommand\sN{{\ensuremath{{\mathcal N}}}}
\newcommand\sO{{\ensuremath{{\mathcal O}}}}

\newcommand\sV{{\mathcal V}}

\newcommand\sR{{\mathcal R}}

\newcommand\sU{{\mathcal U}}

\DeclareMathAlphabet{\pazocal}{OMS}{zplm}{m}{n}

\newcommand{\eps}{{\epsilon}}

\begin{document}

\title{Quantum Theory of Functionally Graded Materials}

\author{Michael J. Landry}
\email{mjlandry@mit.edu}
\affiliation{Quantum Measurement Group, MIT, Cambridge, MA 02139, USA}
\affiliation{Department of Nuclear Science and Engineering, MIT, Cambridge, MA 02139, USA}
\affiliation{Department of Physics, MIT, Cambridge, MA 02139, USA}
\affiliation{Equal contribution to this work}

\author{Ryotaro Okabe}
\affiliation{Quantum Measurement Group, MIT, Cambridge, MA 02139, USA}
\affiliation{Department of Nuclear Science and Engineering, MIT, Cambridge, MA 02139, USA}
\affiliation{Equal contribution to this work}

\author{Chuliang Fu}
\affiliation{Quantum Measurement Group, MIT, Cambridge, MA 02139, USA}
\affiliation{Department of Nuclear Science and Engineering, MIT, Cambridge, MA 02139, USA}

\author{Mingda Li}
\email{mingda@mit.edu}
\affiliation{Quantum Measurement Group, MIT, Cambridge, MA 02139, USA}
\affiliation{Department of Nuclear Science and Engineering, MIT, Cambridge, MA 02139, USA}

\begin{abstract}

Functionally graded materials (FGMs) are composites whose composition or microstructure varies continuously in space, producing position-dependent mechanical and functional properties. In recent years, FGMs have gained significant attention due to advances in additive manufacturing, which enable precise spatial control of composition and orientation. However, their graded, aperiodic structure breaks the assumptions of Bloch’s theorem, making first-principles electronic and electromagnetic calculations challenging. Here we develop an {\it ab initio} quantum theoretical framework for the electromagnetic properties of FGMs. Using a non-interacting electron model, we formulate a theory of modulated Bloch states, derive effective field equations, and solve them by proposing a generalized WKB (GWKB) method, an effective mass approximation, the Boltzmann equation, and numerical approaches. Our GWKB solution is not semiclassical but remains valid in the fully quantum regime. We show that effective observables such as conductivity, magnetic permeability, and electric permittivity generally do not admit a tensorial description in graded media, and that engineered orientational gradients enable precise control of Landau quantization. As a device example, we further develop a theory of graded $p–n$ junctions with enhanced electronic tunability. This framework lays the quantum foundation for predictive design of graded composite materials, enabling AI-accelerated discovery of next-generation functional architectures.

\end{abstract}
\maketitle

\section{Introduction}

Functionally graded materials (FGMs) are composite systems characterized by spatially varying material properties, often realized through gradual changes in composition or microstructure \cite{miyamoto2013functionally,bohidar2014functionally,naebe2016functionally}. Unlike ordinary composites, which typically involve abrupt interfaces between distinct phases, FGMs exhibit continuous gradients that mitigate interfacial stresses and enable novel behaviors that would otherwise be inaccessible in homogeneous media. While such systems have been proposed and experimentally implemented in various engineering contexts, a foundational theoretical framework for understanding their quantum mechanical properties remains underdeveloped. 

To translate this concept into practical applications, manufacturing methods play a significant role. Conventional techniques such as powder metallurgy \cite{jin2005properties,ubeyli2014ballistic}, chemical or physical vapor deposition \cite{sasaki1994thermal,dobrzanski2008structure}, and thermal spraying \cite{stewart2004contact,shanmugavelayutham2007mechanical} have been employed in the fabrication of FGMs. However, these methods face significant challenges, including geometric limitations in terms of part size and structural complexity, constraints on achievable density, high energy consumption, and adverse environmental impact \cite{mahamood2017functionally,reichardt2021advances}. Such drawbacks hindered the broader adoption and advancement of FGMs fabricated via traditional approaches. The recent emergence of additive manufacturing (AM) has been rapidly transforming this field~\cite{zhang2019additive,reichardt2021advances}. Techniques such as directed energy deposition (DED) \cite{singh2021functionally,feenstra2021critical} and powder bed fusion (PBF) \cite{zhang2020additive,ghanavati2025laser} allow precise control over the spatial distribution of materials during fabrication. These techniques make it possible to gradually vary composition, microstructure, or porosity across a part, enabling the production of complex three-dimensional gradients that were previously difficult to achieve. As a result, AM not only makes FGMs more feasible to fabricate but also revitalizes interest in their design, processing, and functional deployment. By overcoming the manufacturing barriers, AM is actively driving the development and renewed exploration of FGMs as a versatile class of advanced materials for structural, thermal, biomedical, and electrical applications.

The potential applications of FGMs span numerous fields \cite{van2025overview}. The design of heterogeneous microstructures in metals and alloys with the help of AM can significantly improve the combination of strength and ductility compared to homogeneous counterparts \cite{chen2024additive}. In aerospace engineering, they offer heat-resistant coatings that withstand extreme thermal gradients, which is also considered the origin of the FGM \cite{koizumi1997fgm}. In thermal management applications, it provides the opportunity to design thermal rectifiers \cite{zhang2025designing} and thermal expansion control to reduce thermal mismatch stress \cite{khor2000effects}. In biomedical implants, FGMs can provide tailored mechanical responses that improve compatibility with bone tissue \cite{watari2004biocompatibility,shi2021functional}. The functional gradient design is also applied in electrochemical devices such as batteries to enable stable cycling with high capacity and better electrochemical performance \cite{zhang2022functionally,jiang2024combinatorial}. In electronic or energy-converting devices, FGMs enable spatially varying band structures and carrier concentration that could enhance charge transport, optical, and energy-converting performance \cite{mueller2003functionally,bharti2013novel,hedegaard2014functionally,kumar2024applications}. Despite great progress on the engineering front, the theoretical tools for describing electromagnetic behavior in FGMs remain underdeveloped. 

Current theoretical approaches often rely on straightforward extensions of traditional solid-state physics techniques. One common such method involves perturbative expansions in which the gradient in material properties is treated as a small deviation from exact periodicity~\cite{LuttingerJ.M.1951TEoa,KohnW.1959APoB,alma990001333990106761}. A common approach that goes beyond perturbation theory involves adding on a slowly-varying potential to an underlying exactly-periodic Bloch potential~\cite{PhysRev.76.1592,alma990002550050106761,SlaterJ.C.1951ASot,LuttingerJ.M.1956QToC,SundaramGanesh1999Wdis}. While effective in limited cases, these methods struggles to capture the full range of FGM behavior, which can include large changes to the periodicity itself. Another common class of approach is to treat the material as a continuous medium, only considering averaged properties~\cite{alma990000628340106761,KohnRobertV.1985Anmf,alma9935461409806761}. While such homogenization methods simplify calculations, they often overlook the role of the microscopic structure of electronic states. Lastly, numerical methods such as finite element analysis provide detailed simulations but can be computationally expensive and may obscure underlying physical principles~\cite{alma990009555520106761,alma990023893840106761,PayneM.C.1992Imtf}. Consequently, there is an urgent need for a robust and versatile analytic framework that can account for the complex and varied properties that characterize the rich field of FGMs. 

This work presents a quantum-mechanical theoretical framework, making use of first and second-quantized methods and the Boltzmann equation to give a comprehensive picture of FGMs and their electromagnetic properties. Treating the electrons as a non-interacting gas subject to a modulated Bloch potential, we formulate a general effective theory for the low-frequency behavior of the system. The resulting equations are simple and can be solved using generalizations of the WKB method and effective mass approximations, or using standard numerical techniques. We discuss easy-to-implement models that connect the parameters of the effective theory to the underlying FGM microstructure and discuss how density functional theory (DFT) can give more precise results in this correspondence, thereby creating a bridge between theory and manufacturing. In particular, we discuss how manufactured orientational order can give precise control over topological features pertaining to Landau levels, even in the absence of a magnetic field. Taking the results from this first-quantized theory, we use second quantization to derive new expressions for the electron, polarization, and current response functions. Lastly, taking the results of the first-quantized theory, we develop an FGM Boltzmann equation, which we use to investigate various material properties and develop a theory of graded diodes. Surprisingly, we find that the effective conductivity, magnetic permeability, and electric permittivity do not admit a tensorial description in FGMs, even when working in the linear regime. Our graded‑diode numerics recover the expected Shockley‑like forward‑bias behavior. Relative to an abrupt junction, grading preserves the built‑in potential while substantially reducing peak fields and spreading the field profile. Exploring alternative grading shapes indicates that deliberate profile design can further reduce field concentration. 

Unlike perturbative or homogenization methods, our approach naturally incorporates strong deviation from periodicity, while maintaining analytic control over observable quantities. Our results reveal that the graded structure of FGMs can be harnessed to enhance electronic and magnetic properties, offering new strategies for tailoring materials to meet specific performance goals. 


\begin{figure*}
  \centering
  \includegraphics[width=1.0\textwidth]{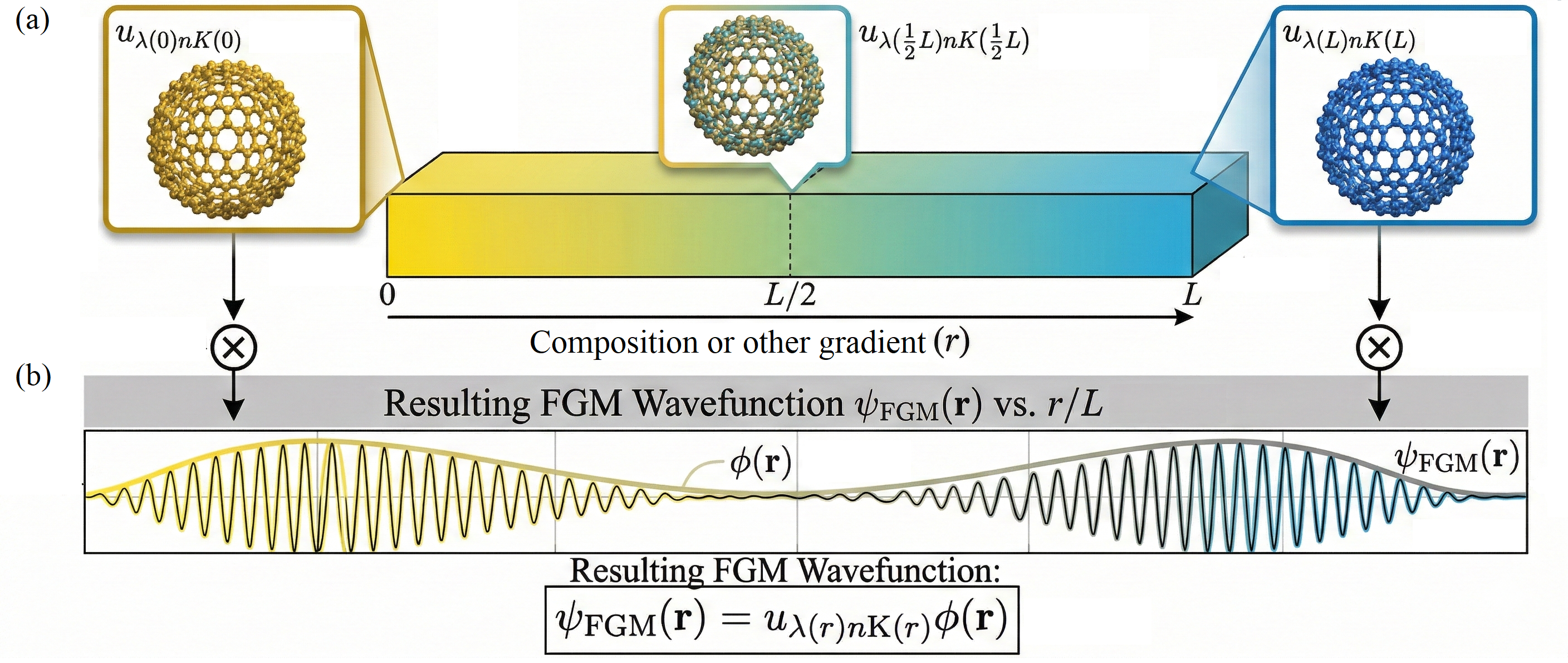}
  \caption{\textbf{Illustration of the wave function ansatz used in the FGM theoretical framework.} (a) The schematic depicts a spatially graded material where the periodicity and lattice parameters vary along the horizontal direction. (b) The wave function is decomposed into a fast-varying modulated Bloch component $u_{\lam(\bm r)nK(\bm r)}(\bm r)$, and a slow-varying envelope function $\phi(\bm r)$, enabling the description of strongly aperiodic systems while retaining analytic control. Perturbative expansion in ratio $\kappa$ of local lattice spacing to lattice-modulation scale allows for fully quantum mechanical effective description valid for strong aperiodicity. Leading-order effective equations for slow modes are given by Eq.~\eqref{eom}. }
  \label{fig1}
\end{figure*}

\section{Theoretical Framework}

There exist many tools for describing the nearly endless array of solid-state materials. When the inter-electron forces can be ignored and the nuclei are arranged in a periodic order, Bloch's theorem dictates that the energy eigenstates take a very particular form, leading to a separated band structure \cite{kittel2018introduction}. By contrast, when the nuclei are arranged in a highly aperiodic order, various well-established mathematical methods can be used to treat such disordered systems~\cite{ANDERSONPHILIPW.2005AoDi,AndersonPhilipW2016Tosg}. But what happens in the middle cases in which the lattice appears nearly periodic on short scales, yet the periodicity can change---sometimes quite drastically---over longer distances? In such cases, both Bloch's theorem and disorder techniques fail. For situations in which the deviation from exact periodicity is sufficiently small, Bloch states paired with standard perturbation theory can give satisfactory results~\cite{alma990012928930106761,LuttingerJ.M.1955MoEa,BelashchenkoK.D.2025GSSS}. But in more extreme cases of lattice modulation, the (approximate) lattice vectors in one location might be substantially different from those elsewhere. As a result, perturbation theory fails. Here new theoretical tools are needed. It is the aim of this section to develop such new tools.

\subsection{Effective theory}

Before moving on to the theory of FGMs, let us briefly review the results of Bloch's theorem. Suppose non-interacting electrons occupy a periodic potential $V(\bm r)$ that is invariant under spatial translations by lattice vectors $\bm R$, that is 
\be
     V(\bm r+\bm R) = V(\bm r).
\ee
Then the resulting energy eigenstates are characterized by two quantum numbers: the band index $n\in \mathbb N$ and the lattice momentum $\bm k$, belonging to the first Brillouin zone, yielding
\be
     \psi_{n\bm k}(\bm r) = u_{n\bm k}(\bm r) e^{i\bm k\cdot \bm r},
\ee
where $u_{n\bm k}(\bm r+\bm R)=u_{n\bm k}(\bm r)$,
such that 
\bega\label{Schru}
     \left[-{1\ov 2m} (\pp+i\bm k)^2 + V(\bm r) - \mE_{n \bm k} \right] u_{n \bm k}(\bm r) = 0, 
\end{gather}
with corresponding eigenenergy $\mE_{n \bm k}$.
Denoting reciprocal lattice vectors by $\bm G$, which are defined such that $e^{i\bm G\cdot\bm R} = 1$ for all lattice vectors $\bm R$, we may expand the Bloch function by
\be\label{expand}
     u_{n \bm k}(\bm r) = \sum_{\bm G} u_{n\bm k,\bm G} e^{i\bm G\cdot\bm r}. 
\ee

Now suppose that we have a parameterized family of periodic potentials $V_\lam(\bm r)$ with lattice vectors $\bm R_\lam$; $\lam$ is some generic parameter or set of parameters. The resulting Bloch wave functions then take the form 
\be
     \psi_{\lam n\bm k}(\bm r) = u_{\lam n \bm k}(\bm r) e^{i\bm k\cdot \bm r},
\ee
with corresponding energy $\mE_{\lam n\bm k}$. 
In the case of FGMs, we want a potential that appears approximately periodic on short distance scales, but whose periodicity can change on longer scales. The most general such potential can be obtained by promoting the parameter $\lam$ to a slowly varying function of position $\lam(\bm r)$. Defining the FGM potential $\sV(\bm r) \equiv V_{\lam(\bm r)}(\bm r)$, we find that the Bloch wave function is no longer a solution to the eigenvalue equation. 
But in the limit that $\lam(\bm r)$ varies on much longer distance scales than the fundamental lattice vectors $\bm R_\lam$, we might expect that a gently modulated Bloch function will yield a good approximation. We therefore suppose the wave function ansatz
\be\label{modBloch}
     \psi_{\rm FGM} (\bm r) = u_{\lam(\bm r) n \bm K(\bm r)}(\bm r) \phi(\bm r),
\ee
where we define $\bm K(\bm r)$ to be the local lattice momentum. A schematic of FGM and the wave functions are shown in Fig.~\ref{fig1}. In principle, there could be coupling between different local Bloch states of the same energy but different $\bm k$ or $n$, in which case our ansatz would need to be modified. However, the couplings between terms of different $\bm k$ will be next-to-leading order in the gradient expansion, while coupling between terms of different $n$ only occur near band crossings. We will leave such considerations for future work, focusing here on the leading-order dynamics for a single, isolated band.  

Letting the subscript PC$(\bm r)$ indicate an integral over a single local primitive unit cell centered at $\bm r$, we postulate the self-consistent relations (see supplementary information) 
\bega\label{Ka}
     \bm K(\bm r)  = \bm k(\bm r) + \bm a(\bm r)   \\ \label{Ka2}
     \bm k(\bm r) = -{i\ov 2} \pp \log \phi(\bm r), \\ \label{Ka3}
     \bm a(\bm r) \equiv \underset {{\rm PC }(\bm r)}{\int} d^3 \bm x \, \bar u_{\lam(\bm r) n \bm K(\bm r)}(\bm x) {1\ov i} \overset\leftrightarrow \pp_{\bm r} u_{\lam(\bm r) n \bm K(\bm r)}(\bm x),
\end{gather}
where we use the notation $f\overset\leftrightarrow \pp g \equiv \ha(g\pp f -f \pp g )$. 
This expression for the local momentum is rather complicated, so its meaning may not be obvious. Do not fear---in the next subsection, we will make physical sense of it. For now, simply notice that in the limit in which $\lam$ is constant and exact periodicity is restored, $\phi\to e^{i\bm k \cdot \bm r}$, we recover the standard Bloch momentum $\bm K(\bm r)\to \bm k$. 

We are interested only in the dynamics of the ``slow'' wave function $\phi(\bm r)$. It is convenient to decompose the wave function into amplitude and phase by $\phi(\bm r) = A(\bm r) e^{i S(\bm r)}$. We assume a separation of scales: the small length scale is the characteristic lattice spacing $\ell_{\rm latt}$, while the large length scale is the modulation scale, namely the characteristic distance over which local lattice spacing changes, $\ell_{\rm mod}$. Letting $\kappa= \ell_{\rm latt}/\ell_{\rm mod} \ll 1$, we work to order $\sO(\kappa^0)$ to obtain leading-order equations for the slow fields. This approximation scheme can be interpreted as a  gradient expansion in which $A$ and $\pp S$ are considered to be zeroth order in derivatives. Implementing a coarse-graining procedure (see supplementary information), we find 
\be\label{eom}\begin{gathered}
    \sE(\bm K;r) =E  ,\quad \pp\cdot (A^2\bm v) = 0,
\end{gathered}\ee
where $\sE_n(\bm K(\bm r);\bm r)=\mE_{\lam(\bm r) n \bm K(\bm r)}$ is the local Bloch energy and $\bm v\equiv {\p\sE\ov\p\bm K } $ is the local electron velocity. Then the above pair of equations in conjunction with the ansatz Eq.~\eqref{modBloch} constitute the leading-order approximation to the FGM wave function in the small $\lam$-gradient expansion. Notice that in assuming the gradients of $\lam$ are small (i.e. $\kappa\ll 1$), we are not assuming small deviations from periodicity. Indeed, the deviations from periodicity may be arbitrarily large on sufficiently large distance scales. 

It is standard in the literature~\cite{HubertM.James1949ESiP} to consider a special subclass of potentials that take the form $V_{\rm Bloch} + V_{\rm slow}$, where $V_{\rm Bloch}$ is an exactly periodic Bloch potential while $V_{\rm slow}$ is a slowly-varying aperiodic potential. By contrast our set-up is far more general, allowing for arbitrary slow modulation of the Bloch potential in three dimensions, thereby characterizing the full range of FGMs.

With Eq.~\eqref{eom} in hand, we have a complete set of equations that govern the slow part of the wave function $\phi=Ae^{iS}$ at leading order in the gradient expansion. In principle, we could stop here and declare that our theory is complete; all that remains is to solve these equations. One could proceed numerically, but we would like to see how far we can get by analytic methods alone. To this end, we will shorty present two approaches that will enable us to find closed-form solutions to these equations: a generalized WBK (GWKB) method and an effective mass approximation. Before solving these equations, however, it is important to understand the nature of the emergent gauge symmetry. 

\subsection{Emergent gauge symmetry}

The term gauge {\it symmetry} is a bit of a misnomer---when constructing physics theories, we ought to think of gauge ``symmetries'' as redundancies of description. There is no actual symmetry present. Any time there are two (or more) mathematical representations of the same physics, we call the transformation that takes us from one description to another a {\it gauge transformation}. With this in mind, let us proceed to discuss our FGM equations.  

The definition of $\bm K(\bm r)$ is rather complicated, mathematically speaking.  Fortunately, however, its  decomposition into $\bm K(\bm r) = \bm k(\bm r) + \bm a(\bm r)$,  admits a simple physical interpretation. 
We can interpret $\bm k(\bm r) = \pp S(\bm r)$ as the local lattice momentum of the slow wave function $\phi(\bm r)$, and $\bm a(\bm r)$ as an emergent gauge field that quantifies how much the local lattice vectors vary, namely the position-space Berry connection. Then $\bm K(\bm r)$ constitutes the {\it local gauge-invariant lattice momentum}.  But why should there be a gauge symmetry for this coarse-grained theory when none exists in the microscopic description? The answer is that we created a gauge redundancy by decomposing the wave function according to Eq.~\eqref{modBloch}. For any scalar function $f(\bm r)$, the full wave function $\psi_{\rm FGM}$---and hence any observable---is invariant under the simultaneous transformation 
\bega
     \phi(\bm r) \to e^{if(\bm r)}\phi(\bm r),\\
     u_{\lam(\bm r) n \bm K(\bm r)}(\bm r) \to e^{-if(\bm r)} u_{\lam(\bm r)n\bm K(\bm r)}(\bm r). 
\end{gather}
At the level of the microscopic wave function, this redundancy of description is utterly trivial: the phases simply cancel. At the level of the coarse-grained equations of motion, however, it appears as a standard (time-independent) $U(1)$ gauge symmetry, with $\bm a(\bm r)$ playing the role of the gauge connection, namely
\be
     \phi(\bm r) \to e^{if(\bm r)}\phi(\bm r),\quad \bm a(\bm r) \to \bm a(\bm r) - \pp f(\bm r). 
\ee
Importantly, this emergent gauge field measures the Berry phase. In general, non-zero Berry curvature arises from band crossings, so when just one band is at play, $\bm a$ is removable by a gauge transformation. If however, a band intersection occurs, then Eq.~\eqref{eom} can be generalized and a non-trivial connection can emerge. The result is a pesudo-magnetic field $\bm b=\pp\times\bm a$. Although our derivation focuses on the single band case, we will investigate some possible consequences of pseudo-magnetic fields in FGM in subsequent sections.

\subsection{Generalized WKB (GWKB) solution} 

We should like to solve the FGM equations for the slow wave function Eq.~\eqref{eom}. Assuming only one band is relevant, we suppose $\bm b=0$, in which case we can gauge-fix $\bm a=0$. Then, supposing that the periodicity of the lattice only changes along the $z$-direction, the equations of motion simplify drastically. In particular, Bloch's theorem now applies so long as we hold $z$ constant. As such, 
\be 
    A(\bm r) = A(z),\quad S(\bm r) = \bm k_\perp\cdot \bm r + \int^z dz' k_z(z'),
\ee
where $\bm k_\perp \equiv (k_x,k_y,0)$ is constant. The equations of motion then reduce to ordinary differential equations,
\be
     \sE = E,\quad (A^2 v_z)' = 0,
\ee
where primes denote derivatives with respect to~$z$. 
These simplified equations invite a generalized, WKB-type approach, which happens to give {\it exact} solutions to these leading-order equations. They are exact in the sense that the gradient expansion truncated at leading order already incorporates the assumption central to the WKB approximation, namely that derivatives of $A$ and $k_z$ are negligible. Our GWKB solution is therefore {\it not} a semi-classical approximation, but is rather a fully quantum mechanical solution given to leading order in the ratio of length scales, $\kappa$. As such, $k_z(z)$ is given implicitly by 
\be
     \sE(\bm k_\perp,k_z(z);z) = E,
\ee
while the modulated amplitude is given by $ A \propto {1/ \sqrt{v_z}}$. 
Normalizing appropriately (see supplementary information), the slow wave function with energy $E$ given by this GWKB solution is then
\bega
     \phi_E(\bm r) = {1\ov \sqrt \mV} \sqrt{\bar v_z \ov v_z(z) } e^{i \bm k_\perp\cdot r + i \int^z dz' k_z(z')},\\ 
     \bar v_z^{-1}\equiv {1\ov \mV}\int d^3\bm r {1\ov v_z(z)},
\end{gather}
where $\mV$ is the volume of the FGM and $\bar v_z$ is the average classical electron velocity. 

There is one important caveat in the above expression. If the wave function crosses into the forbidden region where $k_z(z)$ develops a non-zero imaginary part, then the above expression for the wave function fails. Fortunately, there is a simple trick that can extend any such wave function to the forbidden region: simply analytically continue the energy to the complex plane $E\to E\pm i \eps$ and take the symmetric limit (see supplementary information), 
\be
     \phi_E(\bm r) \propto \ha \lim_{\eps\to 0^+}\left(\phi_{E+i\eps}(\bm r) + \phi_{E-i\eps}(\bm r)\right) . 
\ee
It is well-known that the conventional WKB wave function diverges at the turning points, which separate the classically allowed and forbidden regions. The reason is that at these points, $v_z(z)$ diverges, and the gradient expansion breaks down. This analytic continuation trick can be used to artificially regulate these divergences by promoting $\eps\to\eps(z)$ and keeping $\eps(z)\neq 0$ in the vicinity of any turning points. 

Lastly, we can use this GWKB solution to establish quantization conditions. Placing the FGM in a box of volume $\mV=L^3$, where $L$ is the length of each side, we find that in the transverse directions,
\be
     \bm k_\perp = {\pi \bm n_\perp\ov L},\quad \bm n_\perp\in \mathbb N^2,
\ee
while along the graded direction, in the absence of turning points, we have
\be\label{quant}
     \int_0^L dz\, k_z(z) = \pi n_z,\quad n_z\in \mathbb N.
\ee
We therefore find that, in a finite volume, states are labeled by quantum numbers $\bm n = (\bm n_\perp,n_z)$. If turning points do exist, then the above quantization conditions must be modified. As a simple example, suppose that $a<z<b$ is the classically allowed region and $z<a$, $b<z$ is forbidden, then
\be\label{quant2}
     \int_a^b dz\, k_z(z) = \pi n_z + 2 \theta_0,\quad n_z\in \mathbb N,
\ee
while if $0<z<a$ is allowed and $a<z<L$ is forbidden, then 
\be\label{quant3}
     \int_0^a dz\, k_z(z) = \pi n_z + \theta_0,\quad n_z\in \mathbb N,
\ee
where $\theta_0$ is some constant phase factor; in the case of the conventional WKB approximation, $\theta_0 = \pi/4$~\cite{alma990026400130106761}. We will typically be concerned with situations in which the components of $\bm n$ are large, so $\theta_0$ can be neglected.

\subsection{Effective mass approximation}

Sometimes dealing with the full local Bloch energy $\sE(\bm K(\bm r);\bm r)$ is too cumbersome and we wish to simplify our calculations. In these circumstances, the effective mass approximation can be very useful. Typically this approximation assumes that the local lattice momentum $\bm K(\bm r)$ is small or at least close to some local energy minimum or maximum at $\bm K_0(\bm r)$. 
Expanding to quadratic order, the most general expression for the local Bloch energy becomes
\be\label{sE}
     \sE = U(\bm r) + \ha \delta\bm K(\bm r) \cdot \mathbb M^{-1}(\bm r) \cdot \delta\bm K(\bm r) + \sO(\delta k^4),
\ee
where 
$\delta \bm K(\bm r) \equiv \bm K(\bm r) - \bm K_0(\bm r)$, $U(\bm r)$ is some slow function of $\bm r$, and $\mathbb M(\bm r)$ is the slowly-varying effective mass matrix. Oftentimes, it is convenient to approximate the effective mass by a scalar $\mathbb M(\bm r) \to m^*(\bm r)$. It is understood that $U$ and $\mathbb M$ are implicit functions of both band index $n$ and local extremum point $\bm K_0$. We will typically absorb $\bm K_0$ into the gauge field, defining the new effective gauge field by 
\be\label{ta}
    \tilde {\bm a}(\bm r) \equiv \bm a(\bm r) - \bm K_0(\bm r). 
\ee 
Thus a pseudo-magnetic field arises if $\pp\times \bm K_0\neq 0$. In subsequent discussions, we will simply denote the effective pseudo gauge connection by $\bm a$, leaving the possibility that it includes a contribution from $\bm K_0$ as understood.

Now suppose that the mass matrix $\mathbb M(\bm r)$ varies slowly so that we may neglect its second derivative. Then the slow equations of motion take the convenient form 
\be
     \left[-\ha \bm D_{\bm a}\cdot \left(\mathbb M^{-1} (\bm r) \cdot \bm D_{\bm a} \right) + U(\bm r) - E \right]\phi = 0, 
\ee
where we define the covariant derivative
\be
     \bm D_{\bm a}\,\phi\equiv (\pp+i\bm a) \phi.
\ee
This equation looks very much like the standard envelope function approximation except with the addition of the gauge field $\bm a$~\cite{HubertM.James1949ESiP}. Crucially, however, it was derived with much less restrictive assumptions. In particular, our derivation allows generic, slow deviations from periodicity in the underlying lattice. These deviations may be arbitrarily large.

Now let us suppose that the only gradients of the FGM are along the $z$-direction, that $\bm a=0$, and that the effective mass is well-approximated by the scalar $m^*(z)$. Then we can proceed with the GWKB solution. The simplified form of $\sE$ allows for a simple expression for the local lattice momentum
\be\label{kz}
     k_z(z) = \pm \sqrt{2m^*(z) \left( E - U(z) - {\bm k_\perp^2\ov 2m^*(z)} \right)},
\ee
where $\bm k_\perp=(k_x,k_y,0)$ is constant. Supposing that no turning points are encountered, the slow envelope function for a state with energy $E$ is
\be\label{phiE}
     \varphi_E(\bm r) = {1\ov \sqrt\mV} \sqrt{\bar v_z\ov v_z(z)} e^{i\bm k_\perp\cdot \bm r + i \int^z dz' k_z(z') },
\ee
with $v_z(z) = k_z(z)/m^*(z)$.

\subsection{Pseudomagnetic fields in the effective mass approximation}

\begin{figure*}
  \centering
  \includegraphics[width=1\textwidth]{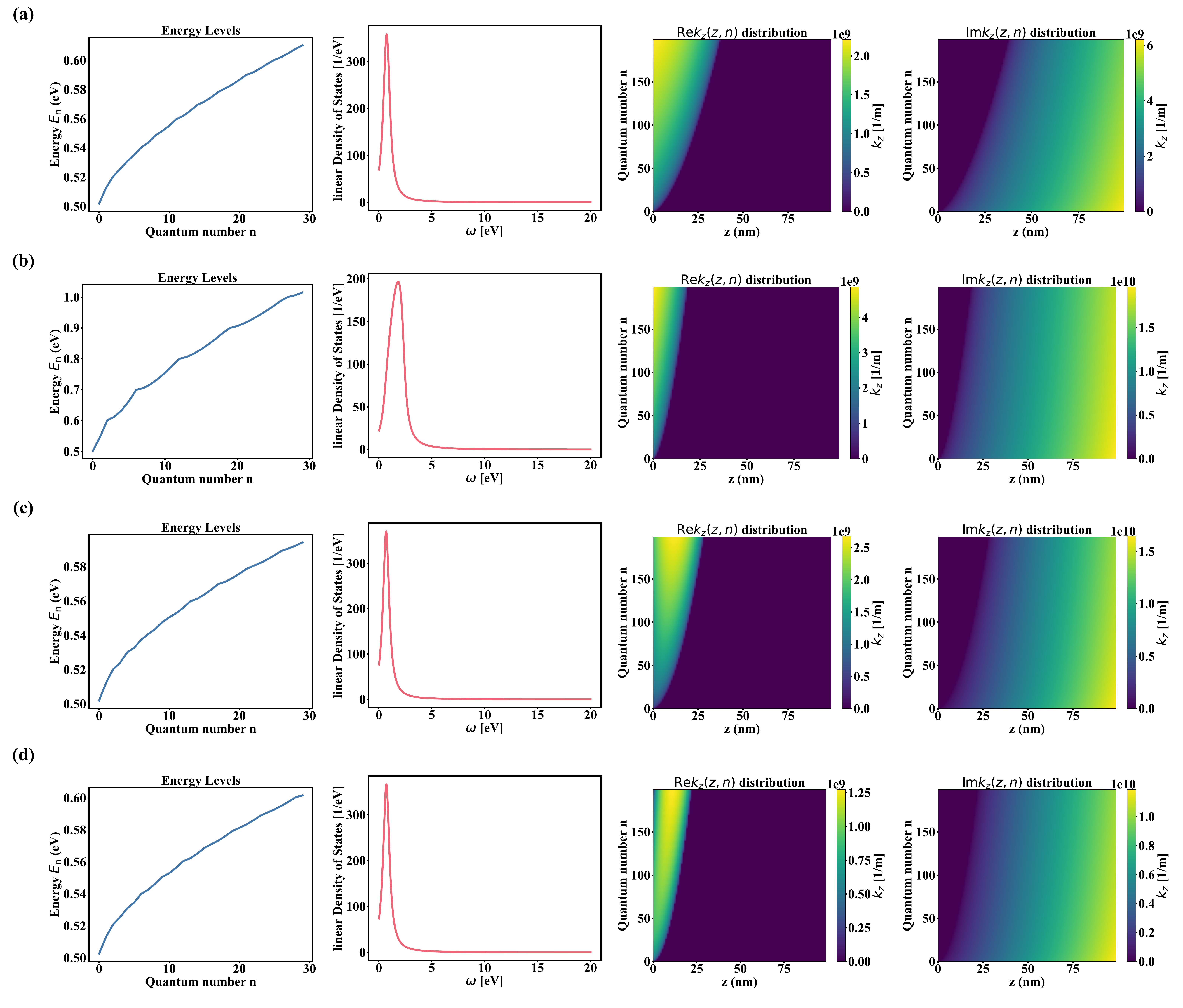}
  \caption{\textbf{Calculated energy spectrum, linear density of state, and $k_z$ from the heuristic models of different FGM system.} From left to right, the energy spectrum, the linear density of state, real part and imaginary part of $k_z$ are presented from the first column to the fourth column. \textbf{(a)}-\textbf{(d)} represent the cases for four different heuristic FGM systems. The system is discretized along the $z$-direction with a spatial step size of $dz = 1\times 10^{-8}\,\mathrm{m}$, resulting in a total of $100$ grid points over a length of $1\,\mu\mathrm{m}$. For \textbf{(a)}, we have $U(z) = (0.5 + 0.01z)$eV, $m^*=(0.5 + 0.01z)$m$_{\mathrm{e}}$, $b=10^{12}$m$^{-2}$. For \textbf{(b)}, we have $U(z) = (0.5 + 0.1z)$eV, $m^*=(0.5 + 0.01z)$m$_{\mathrm{e}}$, $b=10^{12}$m$^{-2}$. For \textbf{(c)}, we have $U(z) = (0.5 + 0.01z)$eV, $m^*=(0.5 + 0.1z)$m$_{\mathrm{e}}$, $b=10^{12}$m$^{-2}$. For \textbf{(d)}, we have $U(z) = (0.5 + 0.01z)$eV, $m^*=(0.5 + 0.05z)$m$_{\mathrm{e}}$, $b=10^{16}$m$^{-2}$. Here m$_{\mathrm{e}}$ reresents the electron mass.}
  \label{result1}
\end{figure*}

We have now introduced two powerful tools for describing FGMs: a GWKB solution and an effective mass approximation. Although our equations have been derived for single-band dynamics, we should like to give the reader a taste of how these mathematical formalisms might treat pseudo-magnetic fields. To keep the math tractable, we will construct a WKB solution using the scalar effective mass approximation and a non-trivial $\bm b=\pp\times \bm a$. We will maintain the assumption that the gradation only exists along the $z$-direction; in particular, suppose $\bm b(\bm r) = \bm b(z)$. This assumption implies
\be
     \bm b(z) = (b_x(z),b_y(z),b_z),
\ee
where $b_z$ is constant. For the case in which $b_z=0$, the emergent gauge field satisfies
\be
     b_x(z)=-\p_z a_y(z),\quad b_y=\p_z a_x(z),\quad a_z=0,
\ee
and we find that the slow wave function takes the form Eq.~\eqref{phiE}, with 
\be\label{kz}
     k_z(z) = \pm \sqrt{2m^*(z) \left( E - U(\bm r) - {\bm (\bm k_\perp+\bm a(z))^2\ov 2m^*(z)} \right)}.
\ee
When $b_z\neq 0$, however, something much more interesting happens: Landau levels emerge. Working in Landau gauge,
\be
     \bm a(\bm r) = (a_x(z),a_y(z)+x b_z,0),
\ee
separate the wave function according to
\be
     \phi(\bm r) = \chi(x,y) \varphi(z). 
\ee
Following standard calculations~\cite{LandauL.D.1958QMNT}, we find that 
\bega
     \chi_{N,\kappa}(x,y) \propto e^{i\kappa y} H_N\left({x+\kappa \ell_{b_z}^2\ov \ell_{b_z}}\right) e^{-(x+\kappa \ell_{b_z}^2)^2/2\ell_{b_z}^2},\\ 
     \ell_{b_z}= 1/\sqrt{b_z},\quad N\in \mathbb N,\quad \kappa\in \mathbb R,
\end{gather}
where $H_N$ are the Hermite polynomials. The corresponding contribution to the local energy is 
\be
     \eps_N(z) = \omega_{b_z}(z)\left(N+\ha\right),\quad \omega_{b_z} = {b_z\ov m^*(z)}. 
\ee
The improved WBK solution yields 
\bega
     \varphi(z) = {1\ov \sqrt\mV} \sqrt{\bar v_z\ov v_z(z)} e^{i\int ^z dz' k_z(z')},\\
     k_z(z) = \pm \sqrt{2m^*(z)\left(E-U(z) - \eps_{N}(z)\right)}. 
\end{gather}
Making use of quantization conditions~\eqref{quant}-\eqref{quant3}, we find that energy $E$ depends on quantum numbers $N$ and $n_z$, but not on $\kappa$. The numerical results for different heuristic FGM systems are presented in Fig.~\ref{result1}. The energy spectrum is calculated with $n_z$ and $N=0$, while $k_z$ is calcuated with $\theta_0 = \pi/4$. The numerical results reveals the sensitivity and tunability of the FGM systems with the spatial gradient. The heuristic profiles used in Fig.~\ref{result1} are generated from Eqs.~\eqref{UHL}--\eqref{SM}.

\subsection{Complete description of FGMs}

Using the effective mass approximation, there are three spatially-varying parameters that must be matched to the physical setup in question. They are the mass $m^*(\bm r)$ (or, more generally, $\mathbb M(\bm r)$) the potential $U(\bm r)$, and the pseudo-magnetic field $\bm b(\bm r) = \pp\times \bm a(\bm r)$. We would like to know in practice how to link these parameters to the physical system. We will consider several cases below:
\begin{itemize}
    \item {\bf Concentration gradients} will typically lead to a non-trivial profile for $U(\bm r)$, as different atomic species contribute different potential profiles. Changes in the electronic structure can also lead to non-trivial profiles for $m^*(\bm r)$. Typically, we should expect $\bm b =0$. Let $X(\bm r)$ parameterize the local concentration such that $X=0$ corresponds to the lowest concentration and $X=1$ the highest. A simple model is to take
    \bega\label{UHL}
          U(\bm r) = (1-X(\bm r)) U_L + X(\bm r) U_H,\\ \label{mHL}
          m^*(\bm r) = (1-X(\bm r)) m^*_L + X(\bm r) m^*_H,
    \end{gather}
    where $U_{H,L}$ and $m^*_{H,L}$ are the potential and effective mass at the highest ($H$) and lowest ($L$) concentrations. 
    \item {\bf Size gradients} can affect the electronic structure, inducing non-trivial profiles for both $U(\bm r)$ and $m^*(\bm r)$, but the pseudo-magnetic field should vanish $\bm b=0$. The model Eqs.~\eqref{UHL}-\eqref{mHL} can be straightforwardly adapted to the case of size gradients. 
    
    \item {\bf Orientational gradients} can generate pseudo-magnetic fields. E.g. in cases in which there is band crossing near the Fermi surface, the dispersion relation becomes linear and $\bm K_0(\bm r)$ should be replaced with the notion of Weyl node separation. It is common to suppose that if the material has orientational order described by $z$-dependent rotations in the $x,y$-plane by angle  $\th(z)$, then~\cite{arjona2018rotational}
    \be\label{eq:K0}
          \bm K_0(\bm r) \sim \theta'(z) (-y,x,0).
    \ee
    In the case of $\theta'=$ const., we find that $\bm b \propto \hat z$ is uniform. 
    If the scalar mass approximation breaks down, then $\mathbb M(\bm r)$ will rotate with the orientational order, while $U(\bm r)$ should remain approximately constant. Letting $\mathbb R(\bm r)$ be the local rotation matrix characterizing orientational order, 
    \bega
          \label{MM}
          \mathbb M(\bm r) =\mathbb R(\bm r) \cdot \mathbb  M_0\cdot \mathbb R^{-1}(\bm r) .
    \end{gather}
    \item {\bf Shape gradients} can induce spatially-varying profiles for both $U(\bm r)$ and $m^*(\bm r)$. The scalar mass approximation may break down, in which case the components of $\mathbb M(\bm r)$ may vary with location. For simple shape gradients that can be specified by a single parameter, models analogous to Eq.~\eqref{UHL} and Eq.\eqref{MM} can be used. In particular, letting $X(\bm r)$ be such a parameter, we then have
    \bega\label{SU}
          U(\bm r) = (1-X(\bm r)) U_L + X(\bm r) U_H ,\\ \label{SM}
          \mathbb M(\bm r) = \mathbb T(X(\bm r)) \cdot\mathbb M_0 \cdot \mathbb T^T(X(\bm r)) ,
    \end{gather}
    where $\mathbb T(X)$ are real matrices. The notion of shape gradients subsumes orientational gradients; as such $\bm b$ can be non-zero, as previously discussed. 
\end{itemize}

The generation of pseudo-magnetic fields in solids is nothing new. Previous works have demonstrated that uniform torsion applied to a crystalline solid can induce Landau levels~\cite{GuineaF.2010Egaa,LevyN2010SPFG,VozmedianoM.A.H.2010Gfig}. Our theory of FGMs, however, provides a novel insight, namely that fabricated orientational order can generate pseudo-magnetic fields. Such order can be manufactured to very precise specifications~\cite{alma990014850120106761,QinDong2010Slfm}---far exceeding the precision and control obtain by applying external stresses to intrinsically periodic materials. To our knowledge, this observation is new.

\section{Correlation functions}

The whole point of this theoretical framework is to compute interesting observables of the FGM. These may include optical, electric and magnetic properties, all of which are ultimately derivable from the electronic response function $G^R(\omega,\bm r,\bm r')$\footnote{While all of these observables are derivable from the electronic response function, we will use more direct means to calculate many of them.}. 
All wave functions must take the form Eq.~\eqref{modBloch}; we will let $\alpha,\beta$ be generic quantum labels to enumerate the FGM states. In this way, we may write $\psi_\alpha(\bm r) = u_\alpha(\bm r) \phi_\alpha(\bm r)$, where $u_\alpha(\bm r)$ are the fast modulated Bloch functions and $\phi_\alpha(\bm r)$ are the slow wave functions. More explicitly, 
\bega
     \psi_\alpha(\bm r) = u_{\lam(\bm r) n \bm K_\alpha(\bm r)} \phi_\alpha(\bm r),
\end{gather}
where $\bm K_\alpha(\bm r)$ is the self-consistently defined gauge-invariant local lattice momentum given by Eq.~\eqref{Ka} and $\phi_\alpha = A_\alpha e^{iS_\alpha}$ satisfies Eq.~\eqref{eom}.

Then the retarded electron Green function is given by
\be\label{elG}
      G^R(\omega,\bm r,\bm r') = \sum_\alpha \psi_\alpha(\bm r) \bar\psi_\alpha(\bm r') (\omega - E_\alpha + 2i/\tau)^{-1},
\ee
where we have introduced by hand a Drude-type relaxation time $\tau$. 

An important physical quantity is the local density of states $\rho(\omega,\bm r) = -{1\ov \pi}\Im G^R(\omega,\bm r,\bm r)$ and its global counterpart $\nu (\omega) \equiv   \int d^3\bm r\, \rho(\omega,\bm r)$. Explicitly, these are given by
\bega
     \rho(\omega,\bm r) ={1\ov \pi} \sum_\alpha |\psi_\alpha(\bm r)|^2 {1/\tau \ov (\omega - E_\alpha)^2 + 1/4\tau^2},\\ \label{dos}
     \nu (\omega) ={1\ov \pi} \sum_\alpha {1/\tau\ov (\omega - E_\alpha)^2 + 1/4\tau^2}. 
\end{gather}
If we are only interested in the slow behavior of $\rho(\omega,\bm r)$, we may coarse-grain over local principle unit cells and obtain an expression that relies only on the slow wave functions, namely
\be\label{eq:rho}
     \rho(\omega,\bm r) = {1\ov \pi} \sum_\alpha |\phi_\alpha(\bm r)|^2 {1/\tau \ov (\omega-E_\alpha)^2 + 1/4\tau^2}. 
\ee
Computing these densities of states requires knowledge of the energy spectrum $E_{\alpha}$. Supposing gradations only along the $z$ direction, the GWBK solution furnishes expressions for $k_z(z)$ in terms of the energy $E$, namely Eq.~\eqref{kz}. The values of $E$ that are permitted must satisfy the quantization conditions Eqs.~\eqref{quant}-\eqref{quant3}, which furnish the energy spectrum.

Lastly, continue to suppose gradations only along the $z$-direction. Then we may label states by the quantum numbers $\bm k_\perp$ and $n_z\in \mathbb N$. We may define the linear density of states by
\be
     g (\omega,\bm k_\perp) = {1\ov \pi } \sum_{n_z} {1/ \tau \ov (\omega- E_{\bm k_\perp,n_z})^2 + 1/4\tau^2}. 
\ee
When a uniform (pseudo-)magnetic field is present in the $z$-direction, we instead have $g(\omega,N)$ in which the energy is now $E_{N,n_z}$. With assumption of $N=0$, the linear density of states are numerically calculated for different heuristic FGM systems in Fig.~\ref{result1}. 

\subsection{Density functional theory}

We will see shortly that many observables can be computed with knowledge of only the slow wave functions $\phi_\alpha(\bm r)$ and energy $\sE_n(\bm K,\bm r)$, but there are other observables that require the full wave functions $\psi_\alpha(\bm r)$. In particular, the electronic green function Eq.~\eqref{elG} inherits fast modes from $\psi_\alpha$ that cannot reasonably be coarse-grained over. Furthermore, when we ultimately compute the current response functions, we will find that interference effects of these wave functions can have important consequences when quantum effects are significant. In such cases, analytic tools are likely to fail and numerical {\it ab initio} techniques like density functional theory (DFT) become necessary to compute the full wave function. Lastly, computing {\it ab initio} expressions for $m^*(\bm r)$, $U(\bm r)$ and $\bm a(\bm r)$  requires solving the microscopic theory.

To compute the full wave functions, we must know $u_\alpha(\bm r)$. In the fully general case, these objects will be very difficult to compute, even with numerical techniques, as the periodicity of the lattice can change quite dramatically from point to point in the sample. Fortunately, however, we do not need to know the exact expression for $u_\alpha$ for the FGM; rather, we can treat each location in the sample as its own distinct exactly periodic lattice and compute the periodic Block function in each region. In particular, hold $\bm r$ fixed and compute the periodic Bloch states as functions of $\bm x$, namely  $u_{\lam(\bm r) n \bm K_\alpha(\bm r)}(\bm x)$. Then these wave functions can be stitched together to give the full wave function by taking $\bm x=\bm r$, yielding $u_{\lam(\bm r) n \bm K_\alpha(\bm r)}(\bm r)$. The full wave function is constructed according to Eq.~\eqref{modBloch} with $\bm K_\alpha(\bm r)$ defined self-consistently by Eqs.~\eqref{Ka}-\eqref{Ka3} and $\phi_\alpha(\bm r)$ given as a solution to Eq.~\eqref{eom}.

Even with the above simplifications, solving for the fast wave functions self-consistently will require substantial computational power. For situations in which the local gauge-invariant lattice momentum $\bm K_\alpha(\bm r)$ is small (as is the assumption in the effective mass approximation), however, the problem simplifies drastically. We can approximate $u_{\lam(\bm r) n \bm K(\bm r)}(\bm x)\approx u_{\lam(\bm r) n \bm K_{0}}(\bm x)$---where $\bm K_{0}$ is the local extremum in Eq.~\eqref{sE}---thereby avoiding any need for solving self-consistent equations. In certain situations the point $\bm K_0$ will remain at a high-symmetry point, so if it is computed in one local Brillouin zone, it may be inferred in others. In cases with reduced symmetry or complicated gradients, there are various techniques that might be employed to compute $\bm K_0$ more efficiently than by brute force. These include coarse DFT calculations with interpolations, and constructing tight-binding models from Wannier functions. 

If this $\bm K\approx \bm K_0$ approximation is inadequate, we can compute corrections via time-independent perturbation theory, by taking Eq.~\eqref{Schru} with $\bm k \to \bm K_0$ to be the unperturbed Hamiltonian and with $\bm k\to \bm K_\alpha$ the exact Hamiltonian.  

\subsection{Second quantization}

We would like to compute the charge and current density response functions, making use of a second quantized formalism. To this end, we will now have a small interlude to explain how response functions in general can be computed in our non-interacting theory. Let $c_\alpha^\dagger$ and $c_\alpha$ be the creation and annihilation operators for electrons in the state $\psi_\alpha$. As electrons are Fermions, these operators satisfy the anti-commutation relations 
\be
     \{c_\alpha^\dagger,c_\beta\} = \delta_{\alpha\beta}. 
\ee 
Now consider two second-quantized operators 
\bega
     A(t) = \sum_{\alpha\beta} A_{\alpha\beta}(t) c_\alpha^\dagger c_\beta,  \\ 
     B(t) = \sum_{\alpha\beta} B_{\alpha\beta}(t) c_\alpha^\dagger c_\beta,
\end{gather}
where $A_{\alpha\beta}(t) = A_{\alpha\beta} e^{-i(E_\alpha-E_\beta)t}$ and $B_{\alpha\beta}(t) =B_{\alpha\beta} e^{-i(E_\alpha-E_\beta)t}$. 
The second-quantized Hamiltonian takes the form $\sH=\sum_\alpha E_\alpha c_\alpha^\dagger c_\alpha$, which allows us to construct the thermal density matrix $\rho\propto e^{-{H-\mu\ov T}}$. The response function for these two operators is defined by
\be
     \Pi^R_{AB}(t-t') \equiv -i\theta(t-t') \langle[A(t),B(t')]\rangle_\rho.
\ee
Making use of standard arguments (see supplementary information),
\be\begin{split}
     \Pi^R_{AB}(t-t') = -i\theta(t-t') \sum_{\alpha\beta} A_{\alpha\beta}(t) B_{\beta\alpha}(t')\\
     \times \left( n_F(E_\beta) -n_F(E_\alpha)\right ),
\end{split}\ee
where $n_F(E) =  2\left[e^{E-\mu}+1\right]^{-1}$, with $\mu=E_F$ the Fermi energy. 
Including dissipative effects amounts to replacing $E_\alpha\to E_\alpha - i/2\tau$ and $E_\beta\to E_\beta+i/2\tau$. Working in frequency domain, we have 
\be\label{response}
     \Pi^R_{AB}(\omega) = \sum_{\alpha\beta} A_{\alpha\beta}B_{\beta\alpha} {n_F(E_\beta + i/2\tau) - n_F(E_\alpha-i/2\tau) \ov \omega - E_\alpha+E_\beta + i/\tau}. 
\ee

\subsection{Polarization and current response functions}

To compute the polarization and current response functions, we work in the second quantized theory. 
The second quantized electron operators are 
\bega
     \Psi(t,\bm r) = \sum_\alpha \psi_\alpha(\bm r) e^{-iE_\alpha t} c_\alpha^\dagger, \\
     \bar\Psi(t,\bm r) = \sum_\alpha \bar \psi_\alpha(\bm r) e^{iE_\alpha t} c_\alpha ,
\end{gather}
where $\psi_\alpha(\bm r)$ are the exact energy eigenstates for the first-quantized theory, with corresponding energy $E_\alpha$.  
The corresponding second quantized Hamiltonian is
\be\label{Hamiltonian}
     \sH = \int d^3 \bm r\, \left[ {1\ov 2m} \pp\bar\Psi\cdot\pp\Psi + \sV \bar\Psi\Psi\right].
\ee
Notice that expanding this Hamiltonian in terms of creation and annihilation operators, we obtain the expected result $\sH = \sum_\alpha E_\alpha c_\alpha^\dagger c_\alpha$. The corresponding action is given by 
\bega
     S = \int dt \int d^3 \bm r  \left [ \bar\Psi i\p_t \Psi -\mH \right] ,
\end{gather}
where $\mH$ is the Hamiltonian density, defined to be the integrand of Eq.~\eqref{Hamiltonian}. To compute the charge and current densities, it is convenient to introduce external (i.e. non-dynamical) electromagnetic sources $A_\mu = (A_0,\bm A)$. In the minimal coupling scheme, we simply replace 
\bega
     \p_\mu\Psi \to D_\mu\Psi \equiv (\p_\mu - iq A_\mu)\Psi,\\
     \p_\mu\bar\Psi \to D_\mu\bar\Psi \equiv (\p_\mu + iq A_\mu)\bar \Psi. 
\end{gather}
where $-q$ is the charge of the electron. The sourced action is then
\be
     S = \int dt\int d^3\bm r \left[\bar\Psi iD_t\Psi - {1\ov 2m} \bm D\bar\Psi\cdot\bm D\Psi -\sV \bar\Psi\Psi \right]. 
\ee
The expressions for the charge and current densities are $J^0(t,\bm r) = {\delta S \ov \delta A_0(t,\bm r)}$ and $\bm J(t,\bm r) = {\delta S \ov \delta \bm A(t,\bm r)}$, that is
\bega
     J^0 = -q \bar\Psi \Psi,\quad \bm J = {iq\ov 2m} \left[\bar\Psi \pp \Psi - \pp\bar\Psi \Psi\right]-{q^2\ov m} \bm A \bar\Psi \Psi. 
\end{gather}

With these second-quantized expressions for the charge and current densities in hand, we can making use of Eq.~\eqref{response} to find the polarization response function 
\be\begin{split}
     \Pi^R_{J^0J^0}(\omega,\bm r,\bm r') = q^2\sum_{\alpha\beta}
     \psi_\alpha(\bm r) \bar\psi_\beta(\bm r) \psi_\beta(\bm r') \bar\psi_\alpha(\bm r') \\
     \times {n_F(E_\beta+i/2\tau) -n_F(E_\alpha-i/2\tau )\ov \omega - E_\alpha+E_\beta + i/\tau}, 
\end{split}\ee
and the current response function
\be\begin{split}\label{current cor}
     \hat\Pi^R_{J^iJ^j}(\omega,\bm r,\bm r') = -{q^2\ov m^2} \sum_{\alpha\beta} 
     \psi_\alpha(\bm r)\overset\leftrightarrow \p_i \bar\psi_\beta(\bm r) \\
     \times \psi_\beta(\bm r') \overset\leftrightarrow \p_j' \bar\psi_\alpha(\bm r') \\
     \times {n_F(E_\beta+i/2\tau) -n_F(E_\alpha-i/2\tau )\ov \omega - E_\alpha+E_\beta + i/\tau}\\
     - {q^2\ov m} \delta_{ij} \delta(\bm r-\bm r')\sum_\alpha n_F(E_\alpha)|\psi_\alpha(\bm r)|^2. 
\end{split}\ee
If quantum interference effects among the fast modes of states $\alpha$ and $\beta$ are negligible, then we can express the polarization and current response functions in terms of the slow wave functions $\phi_\alpha(\bm r)$ by replacing $\psi_\alpha \to \phi_\alpha$ and $\pp\to \bm D_{\bm a}$ in the above expressions. If however, quantum interference effects are important, then a DFT calculation of the fast wave functions becomes necessary. 

The electrical conductivity tensor is given by the Kubo relation
\begin{equation}
    \sigma_{ij}(\omega,\mathbf{r},\mathbf{r}') =  {i\ov  \omega}{\Pi_{J^iJ^j}^{\rm R}(\omega,\mathbf{r},\mathbf{r}')   }.
\end{equation}
This object can be computed numerically once wave functions $\psi_\alpha(\bm r)$ and corresponding energy eigenvalues $E_\alpha$ have been obtained. The result will be a fully quantum mechanical account of the optical conductivity. Often times, however, a semi-classical analysis is sufficient, in which case the calculation of the local optical conductivity tensor greatly simplifies. In the next section, we will see how to compute the local optical conductivity---and many other quantities---in the semiclassical limit using the Boltzmann equation.

\begin{figure*}
  \centering
  \includegraphics[width=0.6\textwidth]{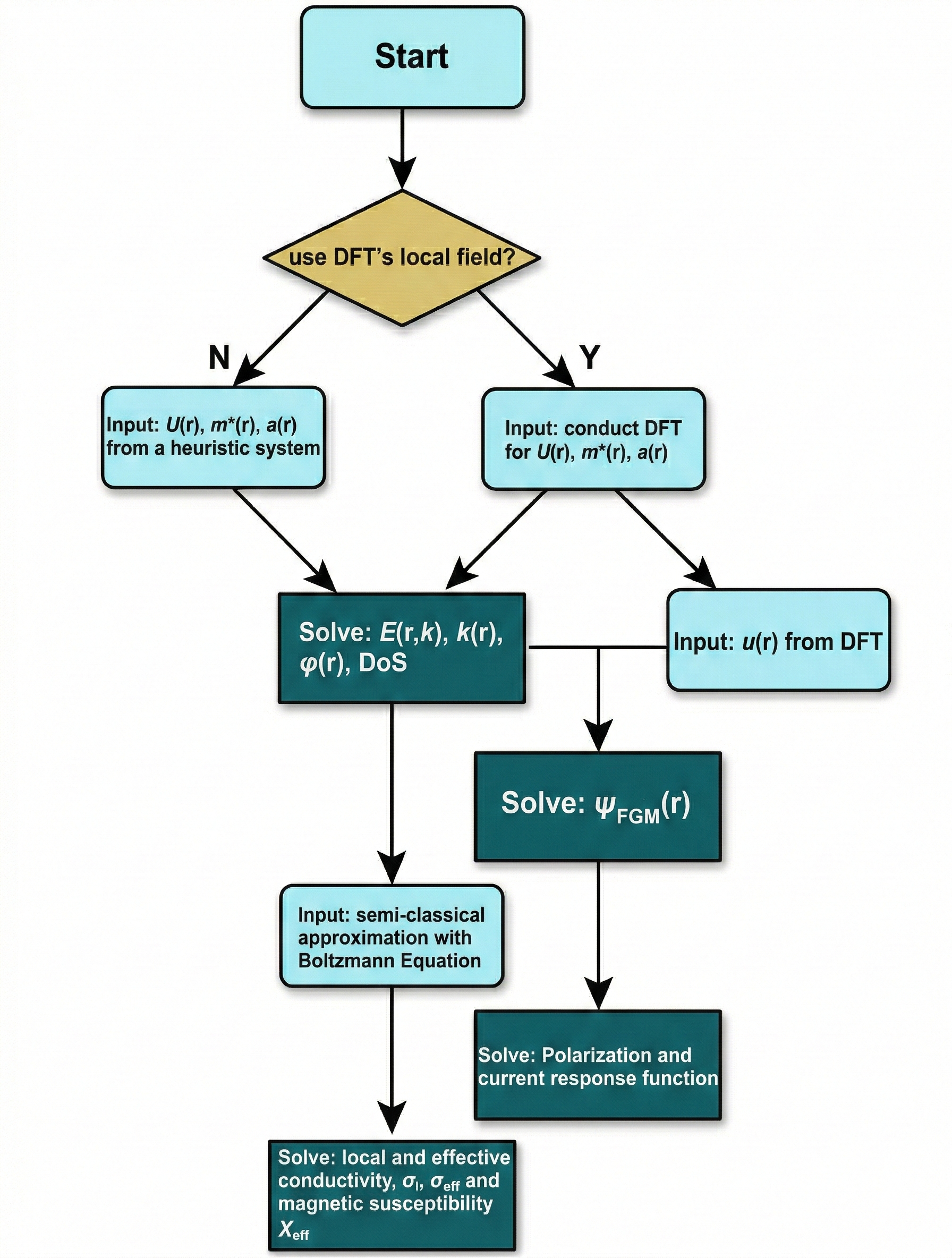}
  \caption{\textbf{Workflow of the theoretical framework for functional graded materials (FGMs).} The flowchart outlines the modeling procedure from inputting physical parameters (from heuristic models or ab initio DFT) to solving for wavefunctions and observables. Depending on the available data and desired precision, one may proceed through different branches, either quantum (via DFT and response theory) or semiclassical (via the Boltzmann equation), to obtain effective transport coefficients such as conductivity and susceptibility.}
  \label{fig2}
\end{figure*}

\noindent The overall workflow of the theory and numerical pipeline is summarized in Fig.~\ref{fig2}.

\section{Boltzmann equation}

\begin{figure*}
  \centering
  \includegraphics[width=1\textwidth]{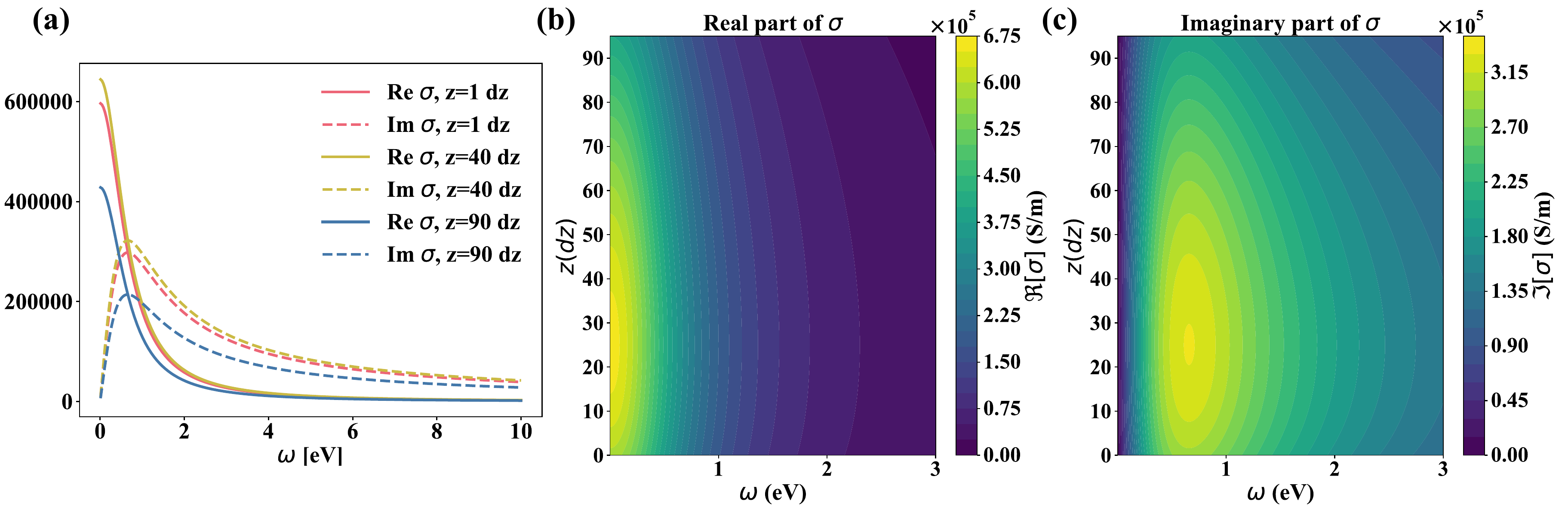}
  \caption{\textbf{Local conductivity calculated for heuristic model.} The assumed heuristic system is discretized along the $z$-direction with a spatial step size of $dz = 1\times 10^{-8}\,\mathrm{m}$, resulting in a total of $100$ grid points over a length of $1\,\mu\mathrm{m}$. For \textbf{(a)}, we have $U(z) = (0.5 + 0.02z)$eV, $m^*=(0.5 + 0.02z)$m$_{\mathrm{e}}$, $b=10^{12}$m$^{-2}$. For \textbf{(a)}, the real part and imaginary part of the local conductivity at different positions are presented. The detailed frequency and z dependent conductivity are also presented in \textbf{(b)} and \textbf{(c)} with its real part and imaginary part.}
  \label{result2}
\end{figure*}

The local conductivity computed from Eq.~\eqref{sigma} using the heuristic profiles in Eqs.~\eqref{UHL}--\eqref{SM} is illustrated in Fig.~\ref{result2}.

In many situations, a semi-classical approximation is valid. Rather than solving the coarse-grained Schrödinger equation in full generality, we can instead use the energy levels $\sE_n$ furnished by Eq.~\eqref{eom} to play the role of a classical Hamiltonian. In particular, take the classical Hamiltonian to be
\be
     H_n(\bm k,\bm r) = \sE_n(\bm k+\bm a(\bm r);\bm r). 
\ee
Let $f_n(\bm k,\bm r;t)$ be the classical distribution function. The Boltzmann equation for the $n^{\rm th}$ band~is
\be
     {\p f_n\ov \p t} + \bm v_n\cdot \pp_{\bm r} f + \bm F_n \cdot \pp_{\bm k} f_n = \left({\p f_n \ov \p t}\right)_{\rm coll},
\ee
where $\bm v_n = \pp_{\bm k} H_n(\bm k,\bm r)$ and $\bm F_n= - \pp_{\bm r} H_n(\bm k,\bm r)$. 

\subsection{Conductivity and current flow}

Supposing that only a single band is relevant, we can neglect the $n$ subscript. We are interested in the linear response of the system to an external electric field $\bm E(\bm r,t)=-\pp\phi(\bm r,t)$. Taking 
\bega\label{f}
     f(\bm k,\bm r;t) = f^0(\bm k,\bm r;t) + \delta f(\bm k,\bm r;t) ,\\ 
     \label{f0}
     f^0(\bm k,\bm r;t) = 2 \left[ e^{H(\bm k,\bm r) - E_F\ov T} +1 \right]^{-1},
\end{gather}
using the relaxation time approximation 
\be
     \left({\p f \ov \p t}\right)_{\rm coll} = -{\delta f\ov \tau},
\ee
and working to linear order in small quantities, we have, in frequency space,
\be
    \delta f(\bm k,\bm r;\omega) =- { q \tau \ov 1-i\omega\tau} \left(-{\p f^0(\bm k,\bm r) \ov \p H}\right) \bm v\cdot \bm E(\bm r,\omega),
\ee
where the electric field is introduced into the Boltzmann equation by replacing $\bm F\to \bm F - q\bm E$. Taking the electric current to be $\bm J = - q\int {d^3\bm k\ov (2\pi)^3} \delta f\bm v$, we can read off the local conductivity tensor 
\be\label{sigma}
     \sigma^{ij}(\bm r,\omega) = { q^2 \tau \ov 1-i\omega\tau} \int {d^3\bm k\ov (2\pi)^3} \left(-{\p f^0\ov \p H}\right) v^i(\bm k,\bm r) v^j(\bm k,\bm r) .
\ee
Notice that as the local conductivity depends on position, if a uniform electric field is applied, then the current is non-uniform, namely $\pp\cdot\bm J\neq 0$. As such, non-zero charge densities can build up. This build-up will eventually break the linear-response approximation.

\begin{figure}
  \centering
  \includegraphics[width=0.48\textwidth]{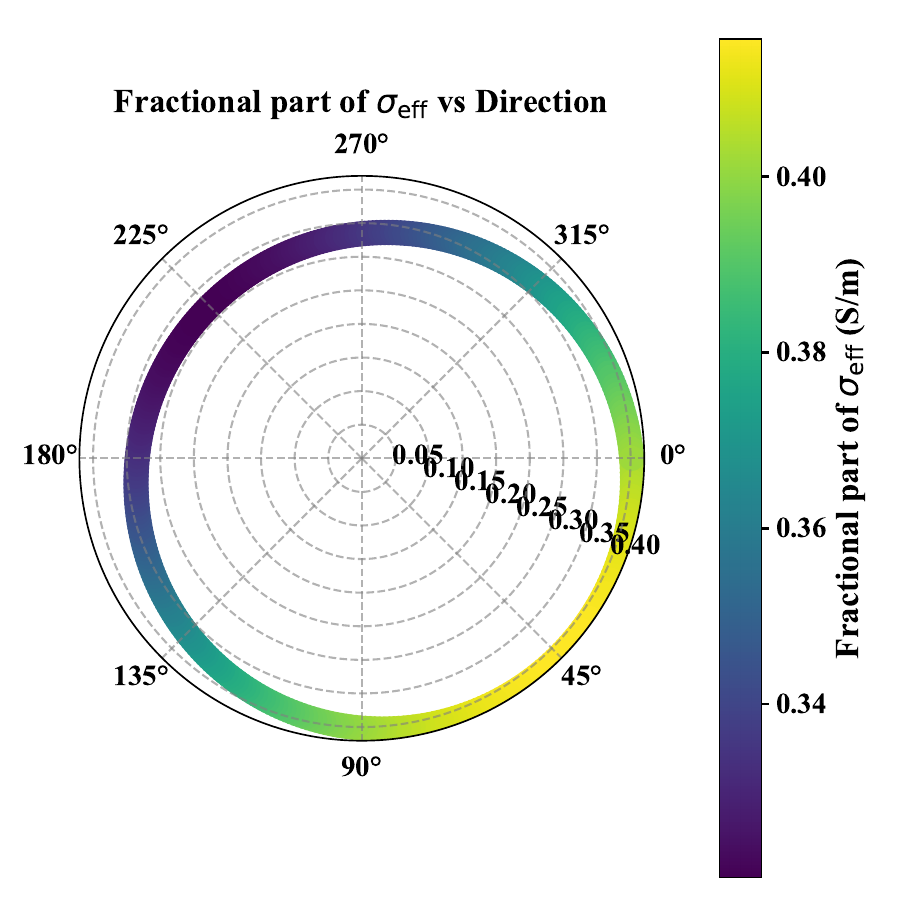}
  \caption{\textbf{Directional dependent effective conductivity.} The calculation is conducted with the corresponding heuristic FGM system with $U(z) = \left[0.5 + 0.1(z\mathrm{cos}\theta +x \mathrm{sin}\theta)\right]$eV, $m^*=\left[0.5 + 0.1(z\mathrm{cos}\theta+x\mathrm{sin}\theta)\right]$m$_{\mathrm{e}}$, $b=10^{12}$m$^{-2}$. As the integer part is identical across all angles, we present only the fractional part to better illustrate the variations. The voltage is applied as 0.5V only through the z direction.  A total of $100$ grid points over a length of $1\,\mu\mathrm{m}$ through both x and z directions. In this relative formulation, the directional dependence of the effective conductivity is effectively captured by fixing the direction of the applied voltage while varying the material gradient direction, which is equivalent to the original setup where the gradient is fixed and the voltage direction is varied.}
  \label{result3}
\end{figure}

To account for effects beyond the linear-response regime, it is helpful to simplify our problem. We suppose that the material has gradations along only one dimension and that the local conductivity tensor can be approximated by a scalar $\sigma^{ij} = \sigma \delta^{ij}$. Apply a potential difference $-V$ along the $z$-direction between the points $z=0$ and $z=L$; we are interested only in the steady-state dynamics, so the divergence of the current vanishes. We thus have the system of equations
\be\label{condeq}
\bm J = -\sigma \pp\phi,\quad \pp\cdot\bm J=0.
\ee 
If the gradations are only along the $z$-direction, we have $J^z = -\sigma(z) \p_z \varphi(z)$, for electrical potential $\varphi$, with boundary conditions $V = \phi(z=0) -\phi(z=L)$. As a result, ignoring edge effects in the $x,y$-plane, the current is
\be\label{series}
    {\sI \ov A_\perp} = {V\ov \int_0^L {dz'\ov \sigma (z')}},
\ee
where $A_\perp=L_\perp^2$ is the cross-sectional area and $L_\perp$ the linear dimension of the graded material perpendicular to the $z$-direction.  
Next, keeping the potential difference along the $z$-direction, suppose now that the material is graded along the $x$-direction. The resulting current density will have a non-trivial $x$-dependence, but we only care about the average current density, so we have
\be\label{parallel}
     {\sI\ov A_\perp} = {V \ov L} \int_0^{L_\perp} {dx'\ov L_\perp} \sigma(x). 
\ee
To arrive at this expression, we must neglect edge effects and suppose that $\bm J \propto \hat z$, which is only guaranteed to hold if $\sigma(x)$ varies slowly. If these conditions do not hold, then we must resort to numerical solutions.  
And lastly, suppose that the material is graded along the direction $\hat w \equiv \hat z \cos\theta + \hat x \sin \theta$ for an arbitrary angle $\theta$. Then the resulting current is (see supplementary information)
\be\label{sI gen}
     {\sI\ov A_\perp} = {V\cos\theta} \int_0^{L\perp} {dx\ov L_\perp} \left[ \int_{x\sin\theta}^{L\cos\theta+x\sin\theta} {dw \ov\sigma(w)} \right]^{-1},
\ee
which holds in the limit that $\sigma(w)$ varies slowly with respect to $x$, namely $\p_w \log\sigma(w) \ll 1/L \sin\theta$. The more general case requires numerically solving Eq.~\eqref{condeq} subject to appropriate boundary conditions. 
It can be checked that taking $\theta\to 0$ reproduces Eq.~\eqref{series} while $\theta\to \pi/2$ reproduces Eq.~\eqref{parallel}. 
From this expression, we define the effective conductivity
\be\label{sigeff}
     \sigma_{\rm eff}(\theta) \equiv {\cos\theta}  \int_0^{L_\perp} {dx\ov L_\perp} \left[ \int_{x\sin\theta }^{L \cos\theta+x\sin\theta} {dw\ov \sigma(w)}  \right]^{-1}. 
\ee
Notice that when the voltage is applied along the direction of gradation, we add the local conductivities in {\it series}, whereas when the voltage is applied perpendicular to the direction of gradation, we add the local conductivities in {\it parallel}. Lastly, notice that we cannot represent the effective conductivity of this material as a tensor as the angular dependence Eq.~\eqref{sI gen} is too general. This non-tonsorial property of the effective conductivity is a unique feature of FGMs that cannot be reproduced by uniform media. The angular dependence from Eq.~\eqref{sigeff} is shown in Fig.~\ref{result3}.

\subsection{Permeability and permittivity}

The non-tensorial nature of the effective conductivity extends to other physical quantities as well. We will see that the effective magnetic permeability and electric permittivity have very similar relationships to their local counter-parts. 

The local magnetization is defined by
\be
     \bm M(\bm r) \equiv - {\mu_B \bm B(\bm r) \ov B(\bm r)}(n_\uparrow(\bm r) - n_\downarrow(\bm r))  ,
\ee
where $n_{\uparrow(\downarrow)}$ are the number densities for spin-up (down) electrons in the presence of magnetic field $\bm B(\bm r)$. The direction of spin is measured with respect to the direction of the magnetic field, so the energy associated with a spin up / down particle is $\pm \mu_B B$.  Working to linear order in the magnetic field, 
\be
     \bm M(\bm r) = \mu_B^2 \bm B(\bm r)  \int {d^3\bm k\ov (2\pi)^3} \left(-{\partial f^0 \ov \partial H}\right),
\ee
where now $f^0(\bm k,\bm r;t) =  \left[ e^{(H(\bm k,\bm r) - E_F)/ T} +1 \right]^{-1}$ differs from the definition Eq.~\eqref{f0} by a factor of two to avoid double counting spin effects.  
The magnetic susceptibility and permeability are therefore 
\bega
     \chi(\bm r) = \mu_B^2 \int {d^3\bm k\ov (2\pi)^3} \left(-{\partial f^0 \ov \partial H}\right) , \\
     \upmu(\bm r) = \upmu_0(1+\chi(\bm r)).
\end{gather}
In the degenerate electron gas approximation, $f^0 = \Th(E_F - H(\bm k,\bm r))$, where $E_F$ is the Fermi energy and $\Th$ the Heaviside step function. Converting the integral over $\bm k$ to an integral over $E$, we have
\be
     \chi(\bm r) = \mu_B^2 \mV \rho(E_F,\bm r) ,
\ee
with $\mV$ the volume of the FGM and $\rho$ is given by Eq.~\eqref{eq:rho}.

The relationship between the $B$-magnetic field and the $H$-magnetic field is given by $\bm B(\bm r)=\upmu(\bm r) \bm H(\bm r)$. Supposing that there are no free currents in the material, $\pp\times\bm H = 0$, so we may introduce the magnetic scalar potential by $\bm H = \pp \psi$. Now consider a slab of material with height $L$ along the $z$-direction that is held at a magnetic potential difference $\sF$, that is $\psi(z=0)=0$ and $\psi(z=L)=\sF$. Notice that the equations of motion are 
\be
     \bm B  = \upmu  \pp\psi  ,\quad \pp\cdot\bm B  = 0. 
\ee
But these equations are identical to the conductivity equations Eq.~\eqref{condeq} with the replacement 
\be
     \bm J \to \bm B,\quad \sigma \to \upmu,\quad \phi\to-\psi,\quad V\to -\sF. 
\ee
Therefore, if $\upmu$ depends only on the coordinate $w=z\cos\theta+x\sin\theta$, then the effective permeability in the limit $\p_w \log\upmu(w) \ll 1/L \sin\theta $ is given in complete analogy to Eq.~\eqref{sigeff}, namely
\be
     \upmu_{\rm eff}(\theta) = {\cos\theta}  \int_0^L {dx\ov L} \left[ \int_{x\sin\theta }^{L \cos\theta+x\sin\theta} {dw\ov \upmu(w)}  \right]^{-1},
\ee
and the effective susceptibility is 
\be \chi_{\rm eff}(\th) \equiv 1-{\upmu_{\rm eff}(\theta)\ov \upmu_0}. \ee 
Thus the effective magnetic susceptibility is non-tensorial, much like the effective conductivity. 

We have a similar story for the electric permittivity $\eps(\bm r)$. In particular, recall that the electric displacement field and the electric field are related by $\bm D(\bm r) = \eps(\bm r) \bm E(\bm r)$. Supposing there are no free charges, the divergence of the electric displacement vanishes, so we have 
\be
     \bm D = -\eps \pp \phi,\quad \pp\cdot \bm D = 0. 
\ee
But these equations are identical to the conductivity equations Eq.~\eqref{condeq} with the replacements 
\be
     \bm J \to \bm D,\quad \sigma\to \eps. 
\ee
As such, the effective permittivity in the limit $\p_w \log\eps(w) \ll 1/L \sin\theta $ is given by
\be
     \eps_{\rm eff}(\theta) = {\cos\theta}  \int_0^L {dx\ov L} \left[ \int_{x\sin\theta }^{L \cos\theta+x\sin\theta} {dw\ov \eps(w)}  \right]^{-1}. 
\ee

The table below summarizes the analogous structures among these non-tensor quantities. 
\begin{center}
\begin{tabular}{ | c || c|| c |}
\hline
\multicolumn{3}{|c|}{Non-tensorial quantities} \\
\hline
 conductivity $\sigma$ & permeability $\upmu$ & permittivity $\eps$ \\ 
 \hline
 $\pp\cdot \bm J=0$ & $\pp\cdot\bm B=0$ & $\pp\cdot \bm D = 0$ \\  
 \hline
 $\bm J = \sigma \bm E$ & $\bm B = \upmu \bm H$ & $\bm D= \eps \bm E$ \\
 \hline
\end{tabular}
\end{center}
We only consider three such quantities here, but in principle, any scalar quantity that adds normally in parallel but as reciprocals in series will exhibit an effective non-tensorial description in FGMs in direct analogy to the conductivity, permeability and permittivity.

\section{Graded diodes}

Diodes are one of the most important examples of graded materials in common use. They involve a sharp junction between $p$-doped and $n$-doped semi-conductors, thereby allowing electric current to only flow in one direction. From the perspective of FGM, however, there is no need to make the $p-n$ junction sharp. Here we will present a theory for gradually-modulated $p-n$ junctions. Numerical solutions to the resulting equations demonstrate that certain kinds of gradations can yield superior performance, most notably an increase in the maximum current the diode can carry. In this section we will make all factors of $k_B$ and $\hbar$ explicit. 

Semiconductors are characterized by two bands: the valence band below the Fermi level and the conduction band above. Let $f_c$ $(f_v)$ represent the distribution function for electrons in the conduction (valence) band and $H_c$ $(H_v)$ its corresponding Hamiltonian. Assuming gradations only along the $z$-direction, defining $f_h\equiv 1-f_v$, and taking the non-degenerate limit, we find
\bega
     f_c(\bm k,z) =  \exp\left\{-{H_c(\bm k,z) - \mu_c \ov k_B T}\right\},\\
     f_h(\bm k,z) =  \exp\left\{{H_v(\bm k,z)-\mu_v\ov T}\right\} ,
\end{gather}
where $\mu_c$ $(\mu_v)$ is the local chemical potential for the conduction (valence) band electrons. Using the scalar mass approximation, 
\bega
     H_c(\bm k,z) = {\hbar^2 k^2\ov 2m_c^*(z)}+ \sU_c(z),\\
     H_v(\bm k,z) = -{\hbar^2 k^2\ov 2m_v^*(z)} + \sU_v(z) .
\end{gather}
Each of these potential energies can be decomposed into the intrinsic potential energy arising from the gradations of the lattice and the electrostatic potential energy
\be
     \sU_c(z) = U_c(z) - q \varphi(z),\quad \sU_v(z)=U_v(z) -q \varphi(z).
\ee
Defining the electron density $n(z)=\int {d^3 k\ov (2\pi)^3} f_c(\bm k,z)$ and hole density $p(z) = \int {d^3 k\ov(2\pi)^3} f_h(\bm k,z)$, we find 
\be 
n(z) = N_c \exp\left\{ {\mu_c+ q\varphi - U_c \ov k_B T} \right\}, 
\ee
and
\be
     p(z) = N_v \exp\left\{ {U_v-\mu_v-q\varphi \ov k_B T} \right\},
\ee
where $N_{c(v)}(z) \equiv 2\left(m_{c(v)}^*(z)k_B T/2\pi\hbar^2\right)^{3/2} $. The electron and hole contributions to the total electrical current $J^z=J^z_n+J^z_p$ are (see Supplementary Information) 
\bega
     J_n^z = q D_n\left(\p_z n+ {n\ov k_B T} \p_z \tilde U_c + {q n \ov k_B T} E_z \right),\\
     \tilde U_c \equiv U_c-k_B T\log N_c,\quad D_c \equiv {\tau k_B T\ov m_c^*},
\end{gather}
and 
\bega
     J^z_p = qD_p\left( -\p_z p + {p\ov k_B T}\p_z \tilde U_v + {qp\ov k_B T} E_z \right),\\
     \tilde U_v\equiv U_v-k_B T\log N_v,\quad D_p\equiv {\tau k_B T\ov m_v^*},
\end{gather}
where $E_z = -\p_z\varphi$ is the electric field. 
There are three equations of motion. The first two describe how electron and hole pairs can recombine, while the third is the electric Gauss law:
\bega
     \p_z J_n^z = q \sR,\quad \p_z J_p^z = -q \sR ,\\
     \p_z^2 \varphi = -{q\ov \varepsilon_0}(p-n+N_d-N_a) ,
\end{gather}
where the rate per unit volume of particle-hole recombination is taken to be
\be
     \sR = {1\ov \tau_{\rm rec}} {np-n_i^2\ov n+p},
\ee
and the intrinsic carrier density $n_i$ is defined by
\bega
     n_i^2 \equiv N_c N_v \exp\left\{-{U_c-U_v\ov k_B T}\right\}. 
\end{gather}
To solve these equations, we must find fix boundary conditions. Suppose the $p$-region is on the left and the $n$-region on the right, such that gradations occur only near $z\sim 0$. Far from the origin, the material parameters are constant (though they may be different constants on the left vs right). Letting $N_a(z)$ represent the acceptor density and $N_d(z)$ the donor density, we suppose that in the $p$-region $N_d(z\to -\infty)=0$, while in the $n$-region $N_a(z\to \infty) = 0$. If each donor and acceptor is fully ionized, and potential bias $V$ is applied, the boundary conditions for the electrical potential are (see Supplementary Information)
\bega
     \varphi(+\infty) = -V+V_{\rm bi},\quad \varphi(-\infty)=0,\\
     V_{\rm bi} \equiv {k_BT\ov q} \log\left\{{N_d(+\infty)N_a(-\infty) \ov N_c(+\infty) N_v(-\infty) e^{- {U_c(+\infty)-U_v(-\infty)\ov k_B T}}}\right\},
\end{gather}
where $V_{\rm bi}$ is the built-in potential difference arising from the material gradations.

To validate the graded diode theory and quantify the advantages of FGM designs, we solve the nonlinear Poisson equation numerically for one-dimensional graded $p-n$ junctions. The detailed numerical methodology (governing equations, discretization, continuation, Newton-Raphson, variable bias, and current density) is given in the Supplementary Information \cite{selberherr1984analysis,markowich2012semiconductor,scharfetter2005large,gummel2005self}. This section presents the results: equilibrium state, I-V characteristics, FGM versus abrupt comparison, and exploration of doping profile shapes \cite{van1950theory,shockley1949theory}.

\subsection{Numerical result for graded diode}

\subsubsection{Simulation setup}

Physical parameters are given in Table~\ref{tab:physical_params}.
\begin{table}[h]
  \centering
  \begin{tabular}{lll}
  \hline
  \textbf{Parameter} & \textbf{Value} & \textbf{Description} \\
  \hline
  $L$ & 10.0 $\mu$m & Device length \\
  $N$ & 401 & Number of grid points \\
  $q$ & 1.0 & Elementary charge (normalized) \\
  $k_B T$ & 0.025 eV & Thermal energy ($\approx$ 300K) \\
  $\epsilon_r$ & 11.9 & Relative permittivity (Si-like) \\
  $\epsilon_0$ & $5.526\times10^{-3}$ & Vacuum permittivity (arb. units) \\
  $\tau$ & 0.01 & Relaxation time (for diffusion) \\
  \hline
  \end{tabular}
  \caption{Physical parameters used in simulation.}
  \label{tab:physical_params}
\end{table}

\begin{figure}[htbp]
  \centering
  \includegraphics[width=0.48\textwidth]{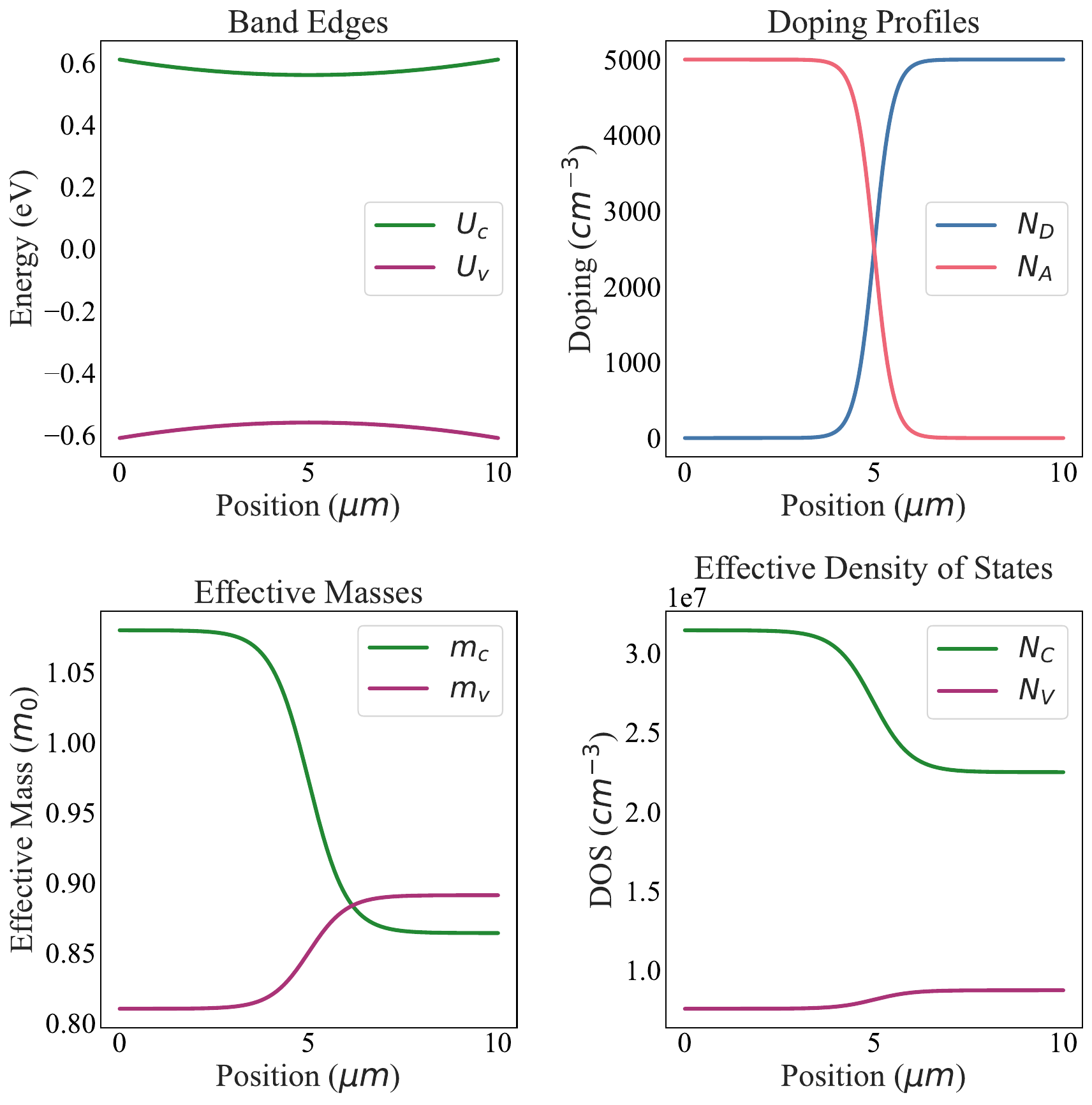}
  \caption{\textbf{FGM material profiles.} (a) Band edges $U_c, U_v$. (b) Doping $N_d, N_a$. (c) Effective masses $m_c, m_v$. (d) Effective DOS $N_C, N_V$.}
  \label{fig_fgm_profile}
\end{figure}

Material parameters are given in Table~\ref{tab:material_params}, and the corresponding spatial profiles are shown in Fig.~\ref{fig_fgm_profile}.

\begin{table}[h]
  \centering
  \begin{tabular}{lll}
  \hline
  \textbf{Parameter} & \textbf{Value} & \textbf{Description} \\
  \hline
  $E_{g,\mathrm{center}}$ & 1.12 eV & Bandgap at center (Si-like) \\
  $\Delta E_g$ & 0.10 eV & Bandgap grading amplitude \\
  $m_c$ range & 1.08 $\to$ 0.864 & Electron mass (p-side $\to$ n-side) \\
  $m_v$ range & 0.81 $\to$ 0.891 & Hole mass (p-side $\to$ n-side) \\
  $N_d$ & $5.0\times10^{3}\,\mu\mathrm{m}^{-3}$ & Donor doping level (n-side) \\
  $N_a$ & $5.0\times10^{3}\,\mu\mathrm{m}^{-3}$ & Acceptor doping level (p-side) \\
  $w_{\mathrm{trans}}$ & 0.5 $\mu$m & Junction transition width \\
  \hline
  \end{tabular}
  \caption{Material parameters for graded junction.}
  \label{tab:material_params}
\end{table}

\begin{figure}[htbp]
  \centering
  \includegraphics[width=0.48\textwidth]{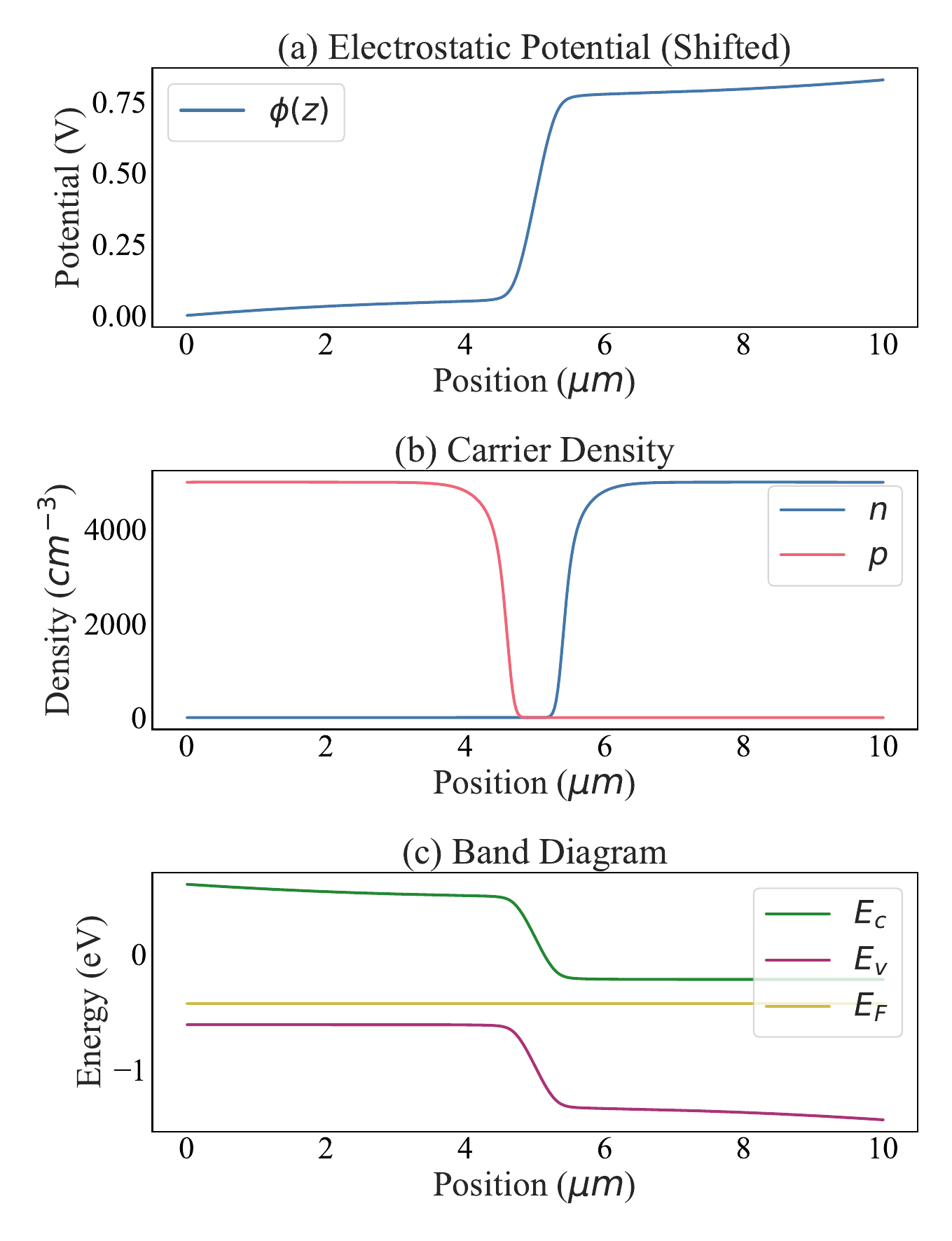}
  \caption{\textbf{Equilibrium solution for the graded diode.} (a) Electrostatic potential $\varphi(z)$. (b) Carrier densities $n(z), p(z)$. (c) Band diagram $E_c(z), E_v(z), E_F$.}
  \label{fig_fgm_num_equilibrium}
\end{figure}

At $V = 0$, the solver gives the equilibrium state (Fig.~\ref{fig_fgm_num_equilibrium}) with the expected built-in potential and band bending \cite{shockley1949theory,van1950theory}.

\subsection{I-V curve for graded diode}

\begin{table}[h]
  \centering
  \begin{tabular}{lll}
  \hline
  \textbf{Parameter} & \textbf{Value} & \textbf{Physical Meaning} \\
  \hline
  $V_{\max}$ & +0.50 V & Forward bias limit \\
  $V_{\min}$ & $-0.50$ V & Reverse bias limit \\
  $\Delta V$ & 0.01 V & Fixed voltage step \\
  \hline
  $V/(k_B T/q)$ & 20 & Normalized voltage (at $V=0.5$ V) \\
  \hline
  \end{tabular}
  \caption{Voltage parameters for I-V curve.}
  \label{tab:voltage_params}
\end{table}

\begin{figure}[htbp]
  \centering
  \includegraphics[width=0.48\textwidth]{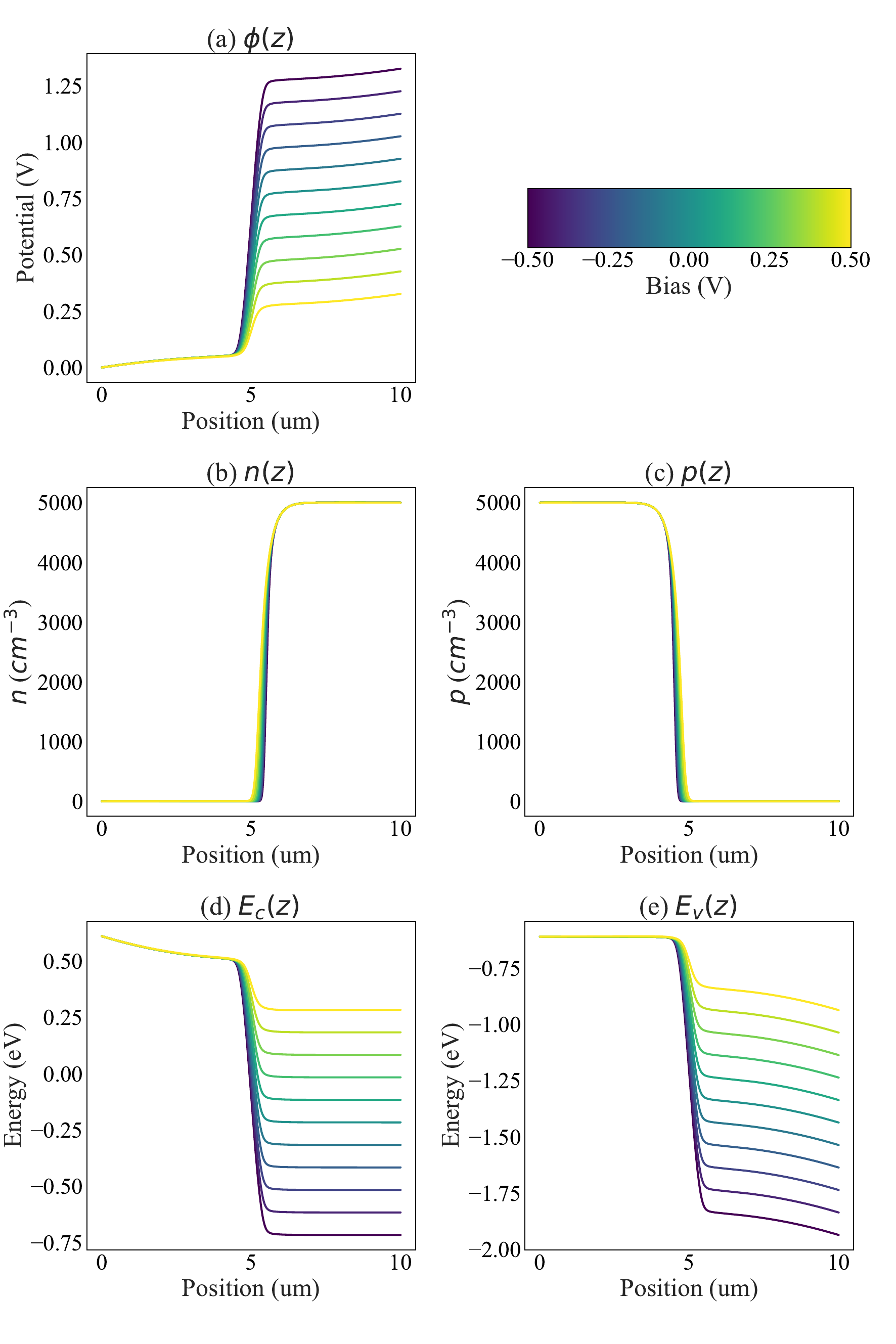}
  \caption{\textbf{Bias-sweep summary (selected voltages).} (a) $\varphi(z)$. (b) $n(z)$. (c) $p(z)$. (d) $E_c(z)$. (e) $E_v(z)$. The color legend indicates the bias values.}
  \label{fig_fgm_num_var_V}
\end{figure}

The plots in Fig.~\ref{fig_fgm_num_var_V} summarize the bias sweep, showing how $\varphi(z)$, $n(z)$, $p(z)$, and the band edges $E_c(z), E_v(z)$ evolve with applied voltage. In this quasi-equilibrium approximation, a single $E_F$ is used \cite{van1950theory}.

\begin{figure}[htbp]
  \centering
  \includegraphics[width=0.48\textwidth]{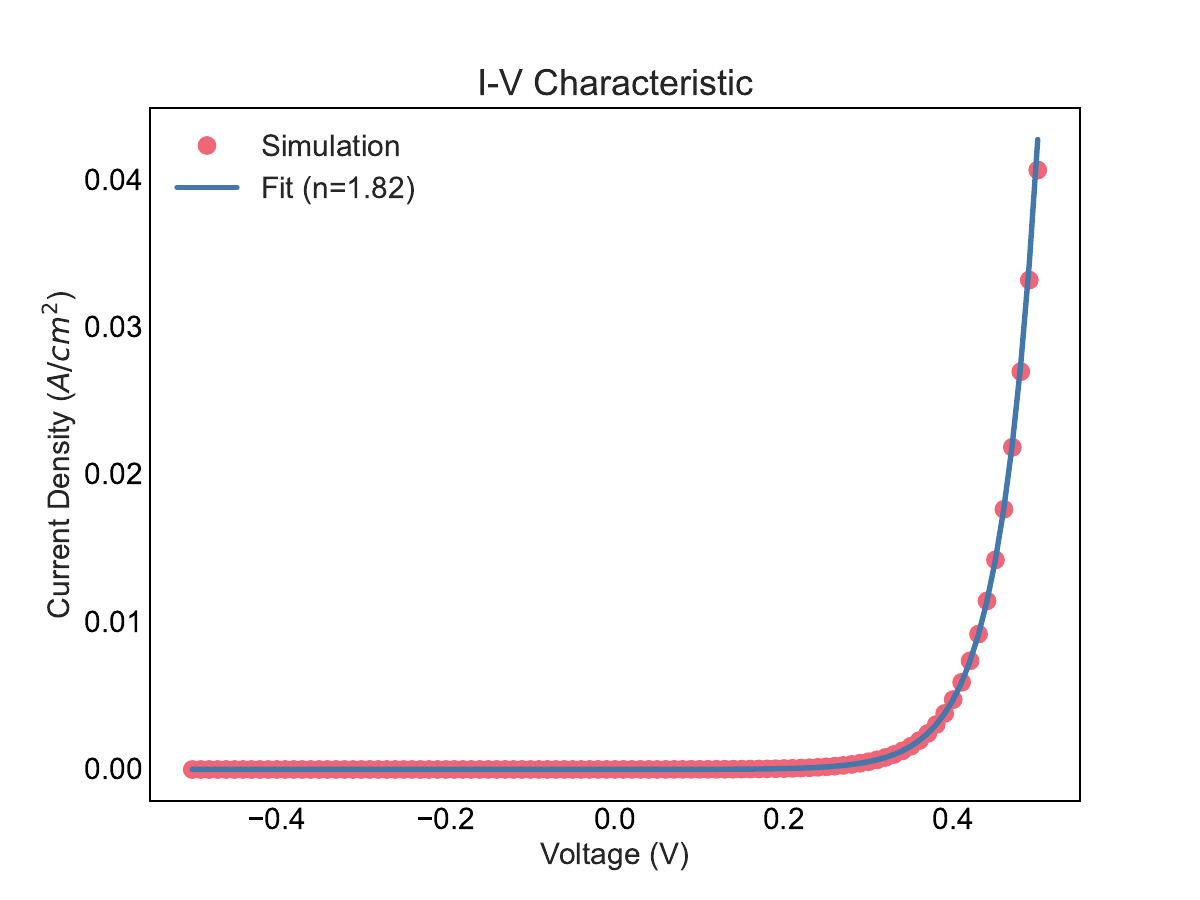}
  \caption{\textbf{I--V characteristic of the graded diode.} Numerical current density as a function of applied bias, with a Shockley fit over the forward-bias region.}
  \label{fig_fgm_iv}
\end{figure}

The calculated I--V curve now follows the expected exponential form in forward bias and can be well fit by the Shockley equation (Fig.~\ref{fig_fgm_iv}). The fit yields $I_s \approx 7.18\times10^{-7}$ and ideality factor $n \approx 1.82$, indicating quasi-equilibrium behavior in this bias range \cite{shockley1949theory,van1950theory,scharfetter2005large,gummel2005self}.

The ideal Shockley diode equation:
\begin{equation}
I = I_0 \left[\exp\left(\frac{qV}{k_B T}\right) - 1\right]
\end{equation}

The plot in Fig.~\ref{fig_fgm_iv} shows the I--V curve for the graded diode together with the Shockley fit over the forward-bias region.

The Shockley equation assumes low-level injection and quasi-equilibrium in the depletion region. In our implementation we use a single Fermi level (quasi-equilibrium), so the model is most reliable in moderate forward bias. With this setup, the simulated I--V curve follows the Shockley form in forward bias and yields a stable fit ($I_s \approx 7.18\times10^{-7}$, $n \approx 1.82$). Extensions such as explicit quasi-Fermi levels $E_{Fn}(z)$, $E_{Fp}(z)$ and full continuity-equation current enforcement can further improve accuracy at higher bias \cite{shockley1949theory,van1950theory,shockley1952statistics,hall1952electron}.

\subsection{Comparison of graded and abrupt diode}

We compare two cases: (1)~FGM: smooth $\tanh$ doping profile; (2)~Abrupt junction: sharper step-like doping. The objective is to quantify how the grading modifies the electrostatic response under identical solver and material settings \cite{shockley1949theory,kroemer1957quasi}.

The key finding is a substantial reduction in peak field: FGM lowers $|E_{\max}|$ by $\sim$44\% and increases the field width (FWHM) relative to the abrupt case (Table~\ref{tab:fgm_abrupt_metrics}). The potential curvature is also reduced, indicating a smoother field distribution.

\begin{table}[h]
  \centering
  \begin{tabular}{lccc}
  \hline
  \textbf{Metric} & \textbf{FGM} & \textbf{Abrupt} & \textbf{$\Delta$} \\
  \hline
  Built-in potential $V_{bi}$ (V) & 0.8266 & 0.8266 & --- \\
  Peak electric field $|E_{\max}|$ (V/$\mu$m) & 1.1982 & 2.1548 & -44.4\% \\
  Field width (FWHM) ($\mu$m) & 0.55 & 0.30 & +83.3\% \\
  Depletion width $W_{\mathrm{dep}}$ ($\mu$m) & 10.0 & 0.90 & --- \\
  Max curvature $|d^2\varphi/dz^2|$ (V/$\mu$m$^2$) & 3.5279 & 7.5956 & -53.6\% \\
  \hline
  \end{tabular}
  \caption{Quantitative comparison of FGM and abrupt junction at equilibrium.}
  \label{tab:fgm_abrupt_metrics}
\end{table}

Overall, the graded profile spreads the electric field over a wider region while keeping the built-in potential unchanged, supporting improved breakdown robustness compared to the abrupt junction.

\begin{figure}[htbp]
  \centering
  \includegraphics[width=0.48\textwidth]{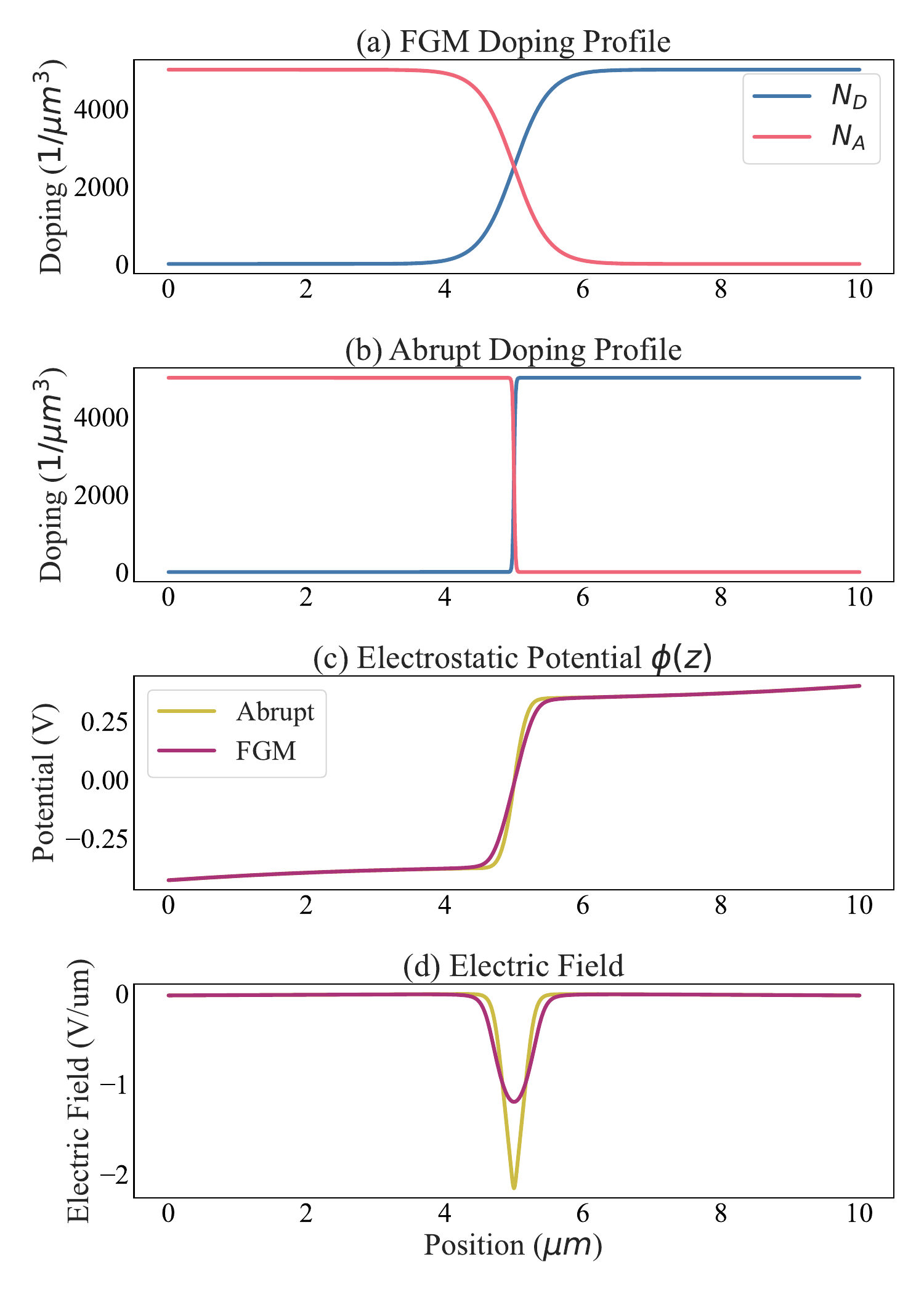}
  \caption{\textbf{FGM vs abrupt junction at equilibrium.} (a) FGM doping. (b) Abrupt doping. (c) Potential $\varphi(z)$. (d) Electric field $E(z)$.}
  \label{fig_fgm_abrupt_comparison}
\end{figure}

\subsection{Exploration of doping profile shapes}

We vary the grading profile to map how shape controls electrostatics under identical material and solver settings (Fig.~\ref{fig_fgm_abrupt_comparison}). The goal is to identify profile families that reduce peak electric field while preserving the built-in potential \cite{kroemer1957quasi}.

\subsubsection{Doping profile types}

\begin{figure}[htbp]
  \centering
  \includegraphics[width=0.48\textwidth]{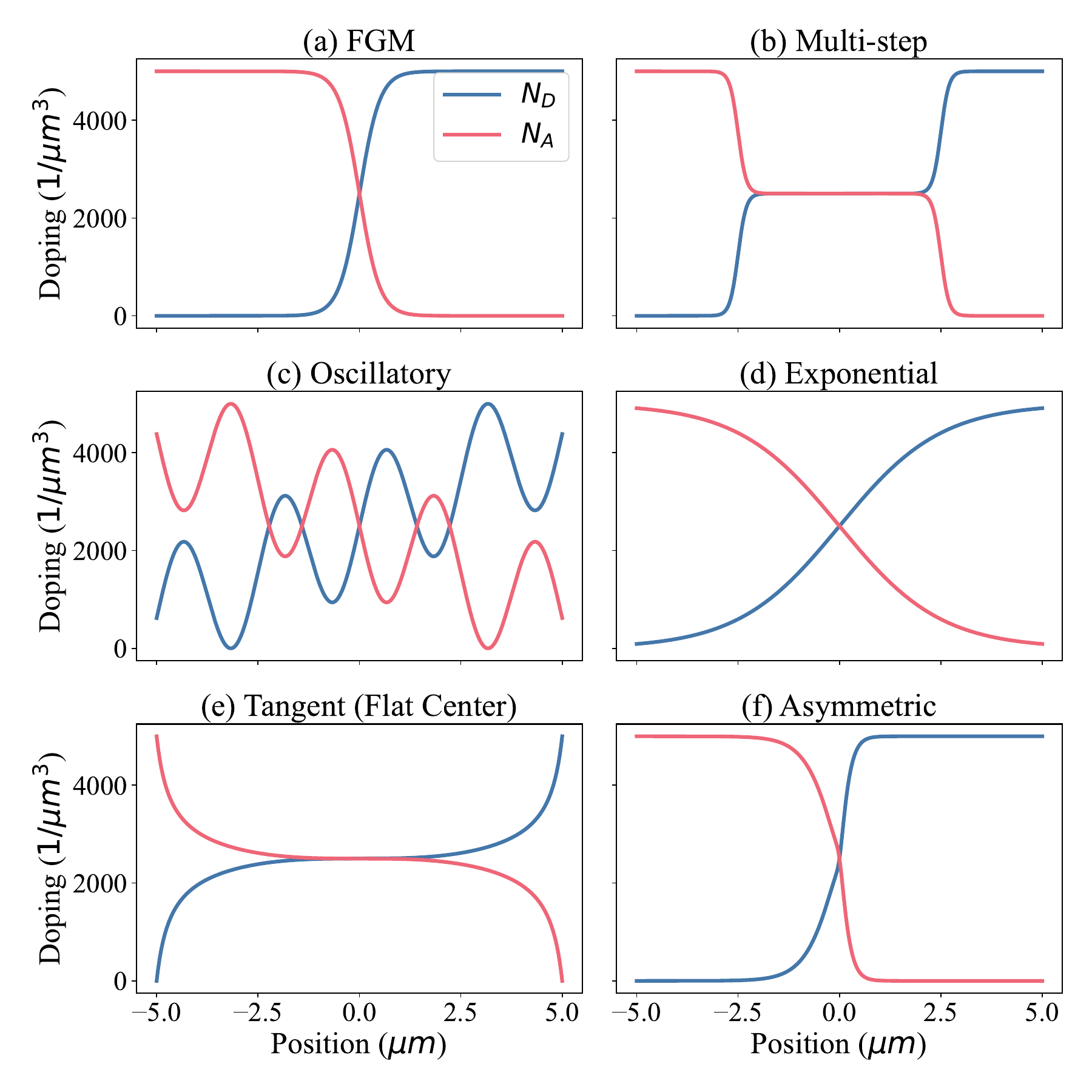}
  \caption{\textbf{Diverse doping profiles explored.} (a) FGM. (b) Multi-step. (c) Oscillatory. (d) Exponential. (e) Tangent. (f) Asymmetric.}
  \label{fig_fgm_diverse_profs}
\end{figure}

We consider six profile types: (1)~FGM (tanh) smooth sigmoidal transition; (2)~Multi-step staged transitions; (3)~Oscillatory sinusoidal modulation; (4)~Exponential concave/convex transition; (5)~Tangent flat at the center with steeper edges; (6)~Asymmetric different transition widths on $p$ and $n$ sides (Fig.~\ref{fig_fgm_diverse_profs}).

\begin{figure}[htbp]
  \centering
  \includegraphics[width=0.48\textwidth]{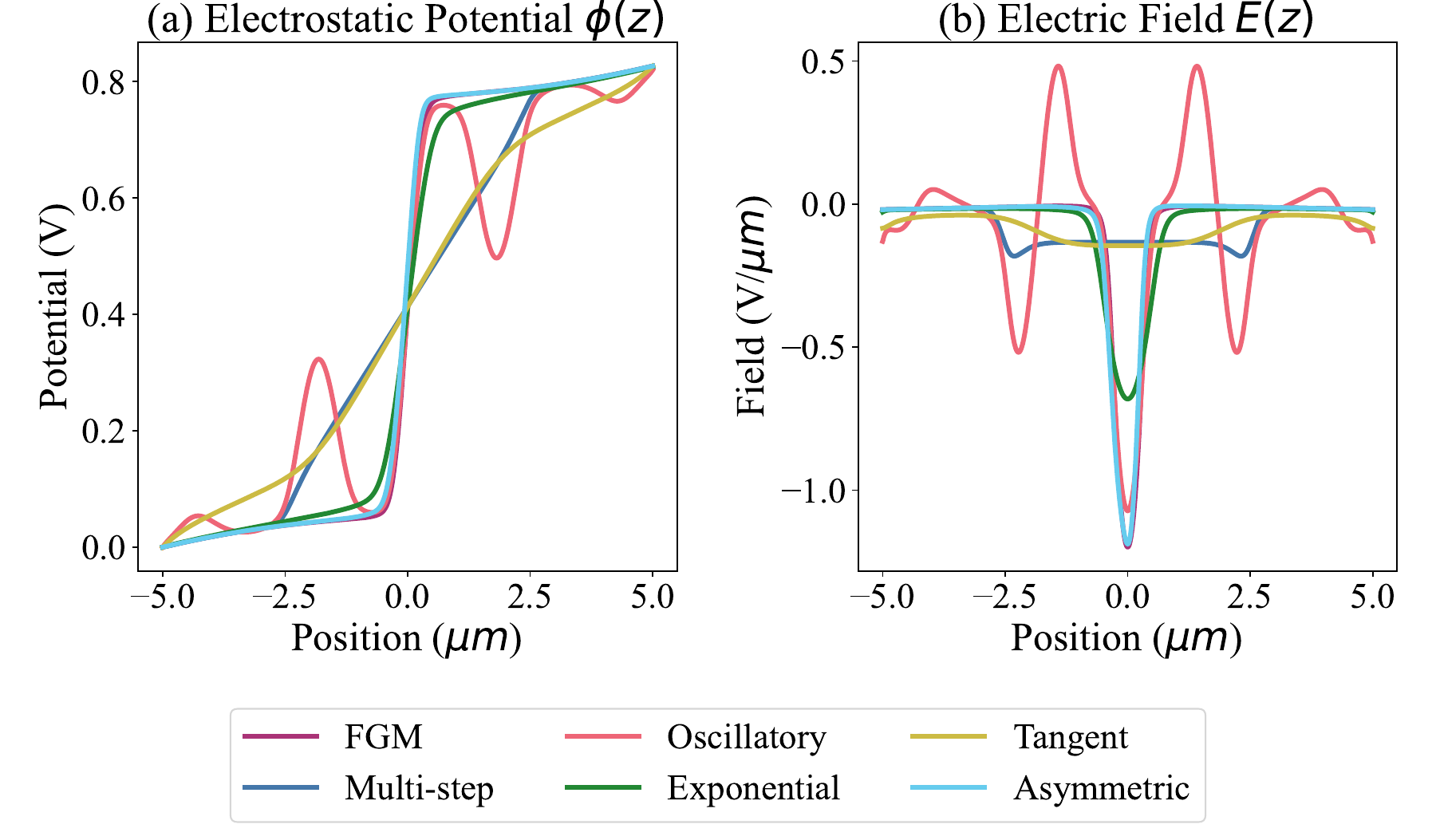}
  \caption{\textbf{Potential and electric field for diverse profiles.} (a) $\varphi(z)$. (b) $E(z)$.}
  \label{fig_fgm_diverse_results}
\end{figure}

Table~\ref{tab:diverse_profiles_metrics} summarizes the quantitative metrics. Two profiles stand out: tangent yields the smallest peak field and curvature, while multi-step also substantially suppresses $|E_{\max}|$. These shapes flatten the central potential drop and spread the field more evenly, making them promising candidates for improved breakdown robustness. Fig.~\ref{fig_fgm_diverse_results} compares $\varphi(z)$ and $E(z)$ across all profiles.


\begin{table}[h]
  \centering
  \begin{tabular}{lcccc}
  \hline
  \textbf{Profile} & $V_{bi}$ (V) & $|E_{\max}|$ & Curvature & Field grad. \\
  \hline
  FGM & 0.8266 & 1.1992 & 3.5646 & 3.5646 \\
  Multi-step & 0.8266 & 0.1824 & 0.5111 & 0.5111 \\
  Oscillatory & 0.8200 & 1.0724 & 2.9042 & 2.9042 \\
  Exponential & 0.8257 & 0.6815 & 1.2359 & 1.2359 \\
  Tangent & 0.8266 & 0.1456 & 0.1015 & 0.1015 \\
  Asymmetric & 0.8266 & 1.1887 & 4.2590 & 4.2590 \\
  \hline
  \end{tabular}
  \caption{Quantitative metrics for diverse doping profile types (units: $V_{bi}$ in V; $|E_{\max}|$ in V/$\mu$m; curvature and field gradient in V/$\mu$m$^2$).}
  \label{tab:diverse_profiles_metrics}
\end{table}

\subsubsection{Summary and conclusions}

Profile shape is a strong design knob. In this study, the tangent and multi-step profiles provide the lowest peak fields, suggesting better field management than FGM, oscillatory, exponential, or asymmetric designs. This motivates targeted profile optimization (e.g., via parametric fitting or automated search) for application-specific constraints.

\section{Conclusion and Future Perspective}

Bloch's theorem represents the first serious attempt to understand the electronic properties of solids with crystalline structures. In an analogous spirit, our theory of modulated Bloch states establishes a quantum foundation for FGMs. Starting from very general considerations, we derive a set of effective equations for non-interacting electrons in a modulated Bloch potential. To solve these equations, we devise two approaches that can be implemented concurrently: a GWKB solution and an effective mass approximation. We emphasize that the GWKB solution is {\it not} semi-classical approximation, distinguishing it conceptually from the standard WKB approximation. Rather, the GWKB  provides the exact solution to the leading-order equations under controlled scale separation between lattice spacing and modulation length. Within the effective mass approximation, any FGM can be characterized completely by a spatially-varying effective mass $m^*(\bm r)$ (or in anisotropic cases the mass tensor $\mathbb M(\bm r)$), an effective potential $U(\bm r)$, and an emergent gauge field $\bm a(\bm r)$. 

To connect these parameters with physical FGM systems, we provide two pathways. First, we introduce heuristic construction rules that map manufactured gradients in composition, size, or orientation onto effective parameters. In particular, engineered orientational order generates controllable pseudo-magnetic fields and tunable Landau quantization, opening practical routes toward manufacturing materials with designed topological or magnetic-like responses, where nontrivial electronic topology is actively programmed through spatially varying orientational order that produces engineered gauge fields and tunable topological spectra. Second, for higher-fidelity modeling, the framework integrates naturally with density functional theory by exploiting locality in graded systems, improving computational efficiency.  

Using second quantization, we derive charge and current response functions and show that quantum interference between fast Bloch components significantly modifies observables in strongly quantum regimes. When such interference is negligible, a Boltzmann formulation yields tractable transport predictions. A central result is that steady-state effective conductivity in FGMs does not generally admit a tensorial description. This departure from conventional tensor behavior arises from the directional dependence introduced by spatial grading and the underlying series–parallel composition of local responses. The same mechanism applies broadly to other response functions, including magnetic susceptibility, dielectric behavior, thermal conductivity, and thermoelectric coefficients. These emergent effective properties are intrinsic to graded media and cannot be replicated in homogeneous or strictly periodic materials, underscoring the fundamentally new response regimes enabled by controlled spatial variation.

As a concrete device example, we formulate and solve graded $p–n$ junctions. Numerical solutions show that smooth grading preserves built-in potentials while redistributing electric fields and increasing allowable current before breakdown. Grading therefore functions not merely as structural variation but as a controllable electronic design parameter.


To our knowledge, this work represents one of the first systematic attempts to address the electronic structure of functionally graded materials from first principles. It establishes a quantum-mechanical foundation for understanding charge transport and field response in spatially graded systems and provides a physically transparent framework that links microscopic structure to macroscopic functionality. This foundation enables rational design of graded composites across a broad technological spectrum, from next-generation microelectronics semiconductor junctions to directionally selective optoelectronic components, high-performing thermal diodes and thermoelectric systems \cite{Li2020ThermalTransportQuantumMaterials}, and even emerging concepts such as optical cloaking and wavefront shaping in architected media.

The importance of this framework is amplified by advances in additive manufacturing, which now permit fine control over compositional, structural, and orientational gradients across multiple length scales. As the accessible design space of graded materials becomes vast, purely empirical or brute-force data-driven exploration becomes inefficient. Our theory provides compact, physics-based descriptors that reduce the effective dimensionality of this space and embed fundamental constraints into the design process. In this way, it enables rational AI-assisted materials discovery \cite{Cheng2026AIDrivenMaterials}, particularly in complex defects and gradient landscapes where unconstrained machine learning alone may struggle to generalize.

Future extensions include incorporating electron interactions, complex disorders \cite{Li2018ElectronPhononDislon}, and phonon transport \cite{Okabe2024VirtualNodeGNN}, enabling a unified treatment of electronic and thermal phenomena in graded media. Integration with DFT workflows and machine-learning-based inverse design strategies will further accelerate predictive optimization of graded architectures, transforming FGMs from empirically engineered systems into systematically designable materials platforms.
\\

\section*{Acknowledgment}
We acknowledge support from the National Science Foundation (NSF) Convergence Accelerator Award No. 2345084, and U.S. Department of Energy (DOE), Basic Energy Sciences (BES), Award No. DE-SC0020148, and the support from R. Wachnik. 

\bibliographystyle{prx}
\bibliography{references}

\clearpage
\onecolumngrid
\setcounter{page}{1}

\vspace{0.5in}

\textbf{\Large Supplementary Information}

\vspace{0.5in}


\normalsize

\makeatletter
\renewcommand \thesection{S\@arabic\c@section}
\renewcommand\thetable{S\@arabic\c@table}
\renewcommand \thefigure{S\@arabic\c@figure}
\renewcommand{\theequation}{S.\arabic{equation}}
\@removefromreset{equation}{section} 
\makeatother
\setcounter{figure}{0}
\setcounter{figure}{0}
\setcounter{table}{0}
\setcounter{section}{0}
\makeatother



\section{Effective theory derivation}

Consider the FGM Hamiltonian 
\be
     H = -{1\ov 2m}\pp^2 +\sV(\bm r),\quad \sV(\bm r) = V_{\lam(\bm r)}(\bm r).
\ee
We are interested in solving the eigenvalue equation $(H-E)\psi=0$ in the limit that gradients of $\lam(\bm r)$ are small. To proceed, consider the action 
\be\label{action1}
     S = -\int d^3\bm r\left[ {1\ov 2m}\pp\bar\psi\cdot\pp\psi + (\sV-E) \bar\psi\psi \right],
\ee
whose equations of motion reproduce the time-independent Schrödinger equation. Decompose $\psi(\bm r) = u(\bm r) \phi(\bm r)$, where $u(\bm r)$ is the fast part of the local Bloch wave function, satisfying 
\bega\label{u eq}
     \left[-{1\ov 2m}(\pp+i\bm K_0)^2 + V_{\lam_0}(\bm r) - \mE_{\lam_0}(\bm K_0) \right] u(\bm r)\bigg|_{\bm r=\bm r_0} = 0,\quad \forall\bm r_0,\\ \label{kr}
     \bm K_0\equiv \bm K(\bm r_0),\quad \lam_0\equiv \lam(\bm r_0),
\end{gather}
where $\bm K(\bm r)$ will be self-consistently defined shortly. 
With this definition of $u$, we may write $u(\bm r) = u_{\lam(\bm r) \bm K(\bm r)}(\bm r)$, where $u_{\lam_0 \bm K_0}(\bm r)$ is a periodic Bloch wave function; choose normalization $\int_{\rm PC} \bar u u =1$. We have suppressed the band index $n$; it is understood that it's value will remain constant throughout this calculation.  

We further decompose the slow wave function into amplitude and phase by $\phi(\bm r) = A(\bm r) e^{i S(\bm r)}$ and substitute into the action~\eqref{action1} to find
\be
     S = - \int d^3\bm r \left\{ {A^2\ov 2}\left[- {1\ov 2m}\bar u (\pp+i\bm k)^2 u + (\sV-E)\bar u u + c.c. \right] + {1\ov 2m}\bar uu(\pp A)^2  \right\}
\ee
where $\bm k(\bm r) \equiv \pp S(\bm r)$. To find the slow equations of motion, we must coarse-grain over fast degrees of freedom. A convenient way to proceed is by averaging the integrand over each local principal cell. Notice that we may express the derivative of $u$ as the sum of two terms
\bega
     \pp u (\bm r) = \left[ \pp_{\bm r} u_{\lam(\tilde{\bm r}) \bm K(\tilde {\bm r})}(\bm r) + \pp_{\tilde{\bm r}} u_{\lam(\tilde{\bm r}) \bm K(\tilde {\bm r})}(\bm r) \right]_{\tilde{\bm r} = \bm r}\\
     \equiv (\hat \pp + \tilde \pp) u(\bm r). 
\end{gather}
As the derivative $\tilde \pp$ only acts on the slow part of $u(\bm r)$, when coarse-graining over a single local principal cell, it can be well-approximated by a slow multiplicative function, which we denote as 
\be
     \tilde \pp  \to i \bm a(\bm r)  ,\quad \bm a(\bm r) \equiv \int_{{\rm PC}(\bm r)}d^3\bm x \, \bar u_{\lam(\bm r) \bm K(\bm r) }(\bm x) {1\ov i} \pp_{\bm r} \bar u_{\lam(\bm r) \bm K(\bm r) }(\bm x) .
\ee
We then define 
\be
     \bm K(\bm r) \equiv \bm k(\bm r) + \bm a(\bm r). 
\ee
As such the action becomes 
\be
     S = - \int d^3\bm r \left\{ {A^2\ov 2m}\left[ -{1\ov 2m}\bar u (\hat \pp+i\bm K)^2 u + (\sV-E)\bar u u + c.c. \right] + {1\ov 2m}\bar uu(\pp A)^2  \right\}.
\ee
Use~\eqref{u eq} to simplify the action 
\be\label{eq:S}
     S=-\int d^3\bm r\left[ {1\ov 2m} (\pp A)^2 + A^2(\sE (\bm K) -E)   \right].
\ee
The corresponding equations of motion are 
\be
   {1\ov 2m} \pp^2 A = A(\sE(\bm K) -E) ,\quad \pp\cdot (A^2\bm v) = 0.
\ee
Notice that l.h.s. of the first equation is higher-order in the gradient expansion, so we may drop it, yielding $\sE(\bm K) = E$.

\section{Normalization of WKB wave function}

Assuming no turning points are encountered, the WKB-type wave function is
\be
     \psi(\bm r) = u(\bm r) {\sN \ov \sqrt{v_z(z)}} e^{i S(z)},
\ee
where $\sN$ is the normalization constant. We have
\be
     1 = \int d^3\bm r |\phi|^2 = \sN^2 \int d^3\bm r|u(\bm r)|^2 {1\ov v_z(z)}. 
\ee
Supposing the fast wave function is normalized by $ \int_{\rm PC} d^3\bm r |u(\bm r)|^2 = 1$ and that $v_z(z)$ can be treated as essentially constant across each local principal cell, we have $\sN^{-1} = \int d^3 \bm r {1\ov v_z(z) } $, and hence 
\be
     \phi(\bm r) = {1\ov \sqrt\mV} \sqrt{\bar v_z\ov v_z(z)} e^{iS(\bm r)},\quad \bar v_z\equiv {1\ov L}\int_0^L {dz v_z(z)}  . 
\ee

\section{WKB and complex analysis}

Consider a 1D Hamiltonian 
\be
     H = -{1\ov 2m} \p_x^2 + \sU(x),
\ee
where $\sU(x)$ is some slowly-varying potential. We wish to solve the eigenvalue problem $(H-E)\psi=0$. Letting $\psi(x) = A(x) e^{iS(x)}$, the equations of motion become
\be
     A'' = A\left[ (S')^2 - p^2 \right],\quad (A^2 S')'=0,
\ee
where $p(x) = \sqrt{2m(E-\sU(x))}$. The WKB approximation amounts to neglecting the second derivative $A''$, yielding 
\be
     S' = \pm p,\quad A\propto {1\ov \sqrt{S'}}. 
\ee
The classically allowed region is defined by the set of $x$ such that $E>\sU(x)$. For the sake of concreteness, suppose that $x<0$ is the classically allowed region, $x>0$ is forbidden and $x=0$ is the turning point. There are two possibilities for the wave function in the classically allowed region, namely
\be
     \psi_{E;\pm}(x<0) = \sN_\pm (E-\sU(x))^{-1/4} e^{\pm i \int^0_x \sqrt{(E-V(x))}}.
\ee
In most textbook treatments, the wave function in the forbidden region is connected to the above wave function via Airy's patch functions; this is a rather cumbersome method. Here, we simply note that if we analytically continue the energy to the complex plane, then we never encounter any turning points. That is $E-\sU(x) = 0$ is only allowed if $E$ is real (assuming that $\sU(x)$ is strictly real). There is, however, an ambiguity in how we might analytically continue: there are two possibilities, $E\to E\pm i \eps$, for small $\eps>0$. We claim that the correct prescription for defining the wave functions of energy $E$ on the full real axis is to take the symmetric limit, namely
\be\label{per}
     \psi_{E;\pm}(x) \equiv \ha\lim_{\eps\to 0^+} \sum_{\eta=\pm 1}  \psi_{E+i\eta\eps;\pm}(x) . 
\ee
The true wave function, which satisfies appropriate boundary conditions is given by some linear combination of these two wave functions $\psi_E(x) = C_+ \psi_{E;+}(x) + C_- \psi_{E;-}(x) $. 
Using the fact that $\lim_{\eps\to 0^+}(-1\pm i\eps)^{-1/4} = e^{\mp {i\pi\ov 4}} $, the wave functions in the forbidden region are
\bega
     \psi_{E;\pm}(x>0) = \ha (\sU-E)^{-1/4} \left[ e^{\pm {i\pi\ov 4}} e^{-\int_0^x dx' \sqrt{2m(\sU(x')-E)}} + e^{\mp {i\pi\ov 4}} e^{+\int_0^x dx' \sqrt{2m(\sU(x')-E)}}  \right]. 
\end{gather}
Imposing the boundary condition $\psi_E(x\to \infty) = 0$, we have $C_\pm = e^{\mp {i\pi\ov 4}} C_0$, for some constant $C_0$. As such the full WKB wave function is
\be
     \psi_E(x) = C_0 \left\{
     \begin{array}{ll}
     2(E-\sU(x) )^{-1/4} \sin \left[ \int_0^x dx' \sqrt{2m(E-\sU(x'))} +{\pi\ov 4} \right] , & x<0 \\
     (\sU(x)-E)^{-1/4} e^{-\int_0^x dx' \sqrt{2m(E-\sU(x))}} , & x>0
     \end{array}
     \right.,
\ee
which agrees with standard results~\cite{alma990026400130106761}.  We therefore see that the general prescription~\eqref{per} is correct.

\section{Computing $\Pi^R_{AB}(\omega)$}

Consider second-quantized operators 
\bega
A(t) = \sum_{\alpha\beta} A_{\alpha\beta }(t) c_\alpha^\dagger c_\beta, \quad
B(t) = \sum_{\alpha\beta} B_{\alpha\beta }(t) c_\alpha^\dagger c_\beta ,
\end{gather}
and thermal density matrix 
\be
     \rho ={1\ov Z} e^{-(\sH-\mu)/T} ,\quad Z = \tr e^{-(\sH-\mu)/T},\quad \sH = \sum_\alpha c^\dagger_\alpha c_\alpha. 
\ee
Compute the expectation value 
\bega
     \vev{ [A(t) ,B(t')]}_\rho = \sum_{\alpha\beta\gamma\delta} \vev{[c_\alpha^\dagger c_\beta ,c_\gamma^\dagger c_\delta] }_\rho e^{-i(E_\alpha-E_\beta)t}e^{-i(E_\gamma-E_\delta)t}  .
\end{gather}
Using the fact that $[c_\alpha^\dagger c_\beta ,c_\gamma^\dagger c_\delta]  = {c_\alpha^\dagger c_\delta}\delta_{\beta\gamma} - {c_\gamma^\dagger c_\beta } \delta_{\delta\alpha} $, we have
\bega
     \sum_{\alpha\beta\gamma\delta} A_{\alpha\beta}(t) B_{\gamma\delta}(t') \left( \vev{c_\alpha^\dagger c_\delta}_\rho \delta_{\beta\gamma} - \vev{c_\gamma^\dagger c_\beta }_\rho \delta_{\delta\alpha} \right) .
\end{gather}
For Fermi-Dirac statistics, $\vev{c^\dagger _\alpha c_\beta}_\rho = \delta_{\alpha\beta} n_F(E_\alpha)$, so 
\bega
     \Pi^R_{AB}(t-t') = -i\th(t-t') \sum_{\alpha\beta} A_{\alpha\beta}(t) B_{\beta\alpha}(t) (n_F(E_\beta)-n_F(E_\alpha)). 
\end{gather}
The Fourier transform is then
\be
     \Pi^R_{AB}(\omega) = \sum_{\alpha\beta} A_{\alpha\beta} B_{\beta\alpha} { n_F(E_\beta)-n_F(E_\alpha)\ov \omega-E_\alpha + E_\beta + i 0^+} . 
\ee

\section{Effective conductivity}

Suppose boundary conditions for the electrical potential $\phi(x,z=0)=V$ and $ \phi(x,z=L)=0$, and let the local conductivity scalar be $\sigma(w)$, for $w=z\cos\theta  +x\sin\theta $. For $z$ held fixed, assume that $\sigma(w)$ varies slowly as a function of $x$, namely $\p_w \log\sigma(w) \ll 1/L \sin\theta $. Then we can suppose the current flows only along the $z$-direction. As such, the steady-state condition $\pp\cdot\bm J=0$ implies $J^z({\bm r}) = J^z(x)$, yielding 
\be
     J^z(x) = - \sigma(w) \partial_z \phi.
\ee
We can then integrate $\p_z \phi$ to find 
\be
     V = J^z(x) \int_0^L {dz\ov \sigma(w)} = {J^z(x) \ov \cos\theta}  \int_{x\sin\theta}^{L\cos\theta+x\sin\theta} {dw \ov\sigma(w)}.
\ee
Define $\sI \equiv \int_0^L dx \int_0^L dy J^z(x) $ and let $A_\perp$ be the cross-sectional area. 
Solving for $J^z(x)$, and averaging over the $x,y$ plane, we then have
\be
     {\sI\ov A_\perp} = {V\cos\theta} \int_0^L {dx\ov L} \left[ \int_{x\sin\theta}^{L\cos\theta+x\sin\theta} {dw \ov\sigma(w)} \right]^{-1}. 
\ee
Notice that as the validity of the approximation is determined by $\p_w \log\sigma(w) \ll 1/L \sin\theta $, there are three distinct cases in which this approximation is valid: (1) $\sigma(w)$ is arbitrary but $\th$ is small (2) $\th$ is arbitrary, but $\sigma(w)$ varies slowly and (3) $\th$ is arbitrary; $\sigma(w)$ may vary quickly but all deviations from its average value are small. If the condition $\p_w \log\sigma(w) \ll 1/L\sin\theta$ does not hold, then we must solve the full system of equations 
\be
     \bm J = - \sigma(w) \pp \phi,\quad \pp\cdot J = 0, 
\ee
subject to the boundary conditions $\phi(x,z=0)=V$ and $ \phi(x,z=L)=0$.

\section{Graded diodes}

The distribution functions for electrons in the conduction band $f_c$ and in the valence band $f_v$ are 
\be
     f_c(\bm k,z) = {1\ov \exp\left\{ H_c(\bm k,z) - \mu_c \ov k_B T \right\} + 1} ,\quad f_v(\bm k,z) = {1\ov \exp\left\{H_v(\bm k,z) - \mu_v \ov k_B T \right\} + 1}
\ee
Define the hole distribution function by $f_h\equiv 1-f_v$. In the non-degenerate limit, we can expand about values of the energy near the energy minimum in the covalence band and near the maximum in the valence band
\be
     f_c(\bm k,z) \to \exp\left\{ -{ H_c(\bm k,z) - \mu_c \ov k_B T} \right\},\quad f_h(\bm k,z) \to \exp\left\{H_v(\bm k,z) - \mu_v \ov k_B T \right\},
\ee
where 
\be
     H_c(\bm k,z) = {\hbar^2 k^2 \ov 2 m_c^*(z)} + \sU_c(z) ,\quad H_v(\bm k,z) = - {\hbar^2 k^2 \ov 2 m_v^*(z)} + \sU_v(z) .
\ee
These potentials can be decomposed into the intrinsic potential energy arising from the gradations of the lattice and the electrostatic potential energy 
\be
     \sU_c(z) = U_c(z) - q\varphi(z),\quad \sU_v(z) = U_v(z) - q \varphi(z) .
\ee
The particle and hole number densities are, respectively, $n(z) = \int {d^3 k\ov (2\pi)^3} f_c(\bm k,z) $ and $p(z) = \int {d^3 k\ov (2\pi)^3} f_h(\bm k,z) $, which are given by direct computation to be
\bega
     n(z) = N_c(z) \exp\left\{ \mu_c(z) + q \varphi(z) - U_c(z) \ov k_B T \right\},\quad p(z) = N_v(z) \exp\left\{ U_v(z) - \mu_v(z) - q\varphi(z) \ov k_B T \right\} ,\\
     N_c(z) \equiv 2 \left( m_c^*(z) k_B T \ov 2 \pi \hbar^2 \right)^{3/2},\quad N_v(z) \equiv 2 \left( m_v^*(z) k_B T \ov 2\pi \hbar^2 \right)^{3/2}
\end{gather}

\subsection{Equilibrium } 
 
It is helpful to define the intrinsic carrier density $n_i$, which is given by $n_i^2 = \bar n \bar p$. The equilibrium values $\bar n$ and $\bar p$ are evaluated at $\mu_c=\mu_v=E_F$, where the Fermi energy $E_F$ is constant. We have
\be
     n_i^2(z) = N_c(z) N_v(z) \exp\left\{-{E_g(z) \ov k_B T}\right\},\quad E_g(z) \equiv U_c(z) - U_v(z) .
\ee
To find the equilibrium boundary conditions, suppose the $p$-region is on the left and the $n$-region is on the right. We suppose that the gradations only occur near the $z\sim 0$ region and that far from this region, the material parameters are constant (though they may be different constants to the right vs the left). Moreover, we will suppose that in the $p$-region, $N_d(z\to-\infty)=0$, while in the $n$-region $N_a(z\to+\infty)=0$. If each donor/ acceptor is fully ionize we may suppose, in the $p$-region,
\be
     \bar p(-\infty) = N_a(-\infty) \implies \bar n(-\infty) = {n_i^2(-\infty) \ov N_a(- \infty)},
\ee
while in the $n$-region
\be
     \bar n(+\infty) = N_d(+\infty) \implies \bar p(+\infty) = {n_i^2(+\infty) \ov N_d(+\infty)}. 
\ee
We can solve these to find 
\be\label{BC}
     \bar \varphi(+\infty) - \bar \varphi(-\infty) = V_{\rm bi}\equiv {k_B T\ov q} \log \left\{ {N_d(+\infty) N_a(-\infty) \ov N_c(+\infty) N_v(-\infty) e^{- {U_c(+\infty) - U_v(-\infty) \ov k_B T}}} \right\}.
\ee
$V_{\rm bi}$ is the built-in potential.

\subsection{Non-equilibrium}

We will now compute what happens in the presence of a voltage bias. We need to take into account non-equilibrium behavior, e.g. how particles and hole pairs can be annihilated. 

The electric current due to particles is (derived in the next subsection) 
\bega
     J_n^z = q D_n \left( \p_z n + {n \ov k_B T} \p_z \tilde U_c + {qn\ov k_B T} E_z \right), \\ 
     \tilde U_c \equiv U_c - k_B T \log N_c,\quad D_n \equiv {\tau k_B T\ov m_c^*} ,\quad E_z = - \p_z \varphi .
\end{gather}
while the electric current due to holes is 
\bega
     J_p^z = q D_p(z) \left(- \p_z p + {p\ov k_B T} \p_z \tilde U_v + {q p \ov k_B T} E_z \right),\\ 
     \tilde U_v \equiv U_v + k_B T \log N_v,\quad D_p \equiv {\tau k_B T\ov m_v^*} . 
\end{gather}

The equations of motion are then 
\be
     \p_z J_n^z = q \sR,\quad \p_z J_p^z = - q \sR,\quad \p_z^2 \varphi = - {q\ov \varepsilon_0} \left( p - n + N_d - N_a  \right) ,
\ee
where the recombination rate can be modeled by
\be
     \sR = {1\ov \tau_{\rm rec}} {n p - n_i^2\ov n+p},\quad n_i^2 \equiv N_c N_v \exp\left\{ - {U_c-U_v\ov k_B T} \right\}. 
\ee

In order to find a solution to these equations, we need to impose the correct boundary conditions. Suppose we want to impose a voltage bias $V$, where the voltage difference between the $n$ region ($z\to +\infty$) and the $p$-region ($z\to-\infty$) is $-V$, we must impose the boundary conditions
\bega
     \varphi(+\infty)  = - V + V_{\rm bi},\quad n(+\infty)=N_d(+\infty),\quad p(+\infty)={n_i^2(+\infty)\ov N_d(+\infty)},\\
     \varphi(-\infty)=0,\quad p(-\infty)=N_a(-\infty),\quad n(-\infty) = {n_i^2(-\infty)\ov N_a(-\infty)}, \\
     V_{\rm bi} \equiv {k_B T\ov q} \log \left\{ {N_d(+\infty) N_a(-\infty) \ov N_c(+\infty) N_v(-\infty) e^{- {U_c(+\infty) - U_v(-\infty) \ov k_B T}}} \right\}. 
\end{gather}

\subsection{Boltzmann equation}\label{A:C}

Consider the time-independent Boltzmann equation in the relaxation time approximation
\be
     {1\ov \hbar}{\p \hat H \ov \p k_z} {\p \ff\ov \p z} - {1\ov \hbar}{\p \hat H \ov \p z}{\p \ff \ov \p k_z} = -{\delta f\ov \tau},
\ee
where $\ff = f + \delta f$. Working to linear order in small quantities allows us to replace $\ff\to f$ on the l.h.s.

First consider electrons in the conduction band, where $f_c = e^{-{H_c-\mu_c\ov k_B T}}$, and $\hat H_c = H_c - q\delta \varphi$, for $\delta\varphi = \varphi-\bar\varphi$. The l.h.s. terms of the Boltzmann equation become
\bega
     {1\ov \hbar} {\p f_c\ov \p H_c} \left[ {\p H \ov \p k_z} \left( {\p H_c \ov \p z} - {\p \mu_c\ov \p z} \right) - \left({\p H_c \ov \p z} - q {\p\delta\varphi\ov \p z} \right) {\p H_c\ov \p k_z}   \right] 
     = {\hbar \ov k_B T m^*_c(z)} k_z f_c \p_z(\mu_c - q\delta\varphi). 
\end{gather}
The resulting electrical current carried by electrons in the conduction band is 
\bega
     J_n^z = -2q \int {d^3 k\ov (2\pi)^3} \delta f_c v_z = {q\tau \ov  m_c^*} n \p_z(\mu_c-q\delta\varphi). 
\end{gather}
Using the identity $n \p_z(\mu_c - q\varphi) = k_B T \p_z n + n \p_z(U_c - q \varphi - k_B T \log N_c)$, letting $E_z = -\p_z\varphi$, we find
\bega
     J_n^z = q D_n \left( \p_z n + {n \ov k_B T} \p_z \tilde U_c + {qn\ov k_B T} E_z \right), \\ 
     \tilde U_c \equiv U_c - k_B T \log N_c,\quad D_n \equiv {\tau k_B T\ov m_c^*} .
\end{gather}

Going though a very similar derivation, the electric current carried by holes is 
\bega
     J_p^z = q D_p(z) \left( - \p_z p + {p\ov k_B T} \p_z \tilde U_v + {q p \ov k_B T} E_z \right),\\ 
     \tilde U_v \equiv U_v - k_B T \log N_v,\quad D_p \equiv {\tau k_B T\ov m_v^*} . 
\end{gather}

\section{Details of Numerical Simulations}

\subsection{Low-temperature limit of local conductivity}

We start from the semiclassical expression for the local conductivity tensor under the relaxation-time approximation:
\begin{equation}
    \sigma^{ij}(z,\omega) = \frac{e^2 \tau}{1 - i \omega \tau} \int \frac{d^3 \bm{k}}{(2\pi)^3}
    \left(-\frac{\partial f^0}{\partial H} \right) v^i(\bm{k}) v^j(\bm{k}),
\end{equation}
where the local dispersion relation is assumed to be parabolic,
\begin{equation}
    H(\bm{k}, z) = \frac{\hbar^2 \bm{k}^2}{2 m^*} + U(z),
\end{equation}
and the velocity is given by
\begin{equation}
    v^i(\bm{k}) = \frac{1}{\hbar} \frac{\partial H}{\partial k_i} = \frac{\hbar k_i}{m^*}.
\end{equation}

In the low-temperature limit, the derivative of the Fermi-Dirac distribution approaches a Dirac delta function:
\begin{equation}
    -\frac{\partial f^0}{\partial H} \xrightarrow{T \to 0} 2\delta\left( H(\bm{k},z) - E_F \right)
    = 2\delta\left( \frac{\hbar^2 k^2}{2 m^*} + U(z) - E_F \right).
\end{equation}

Defining the local Fermi energy as
\begin{equation}
    \epsilon(z) \equiv E_F - U(z), \quad
    k_F(z) = \sqrt{\frac{2 m^* \epsilon(z)}{\hbar^2}},
\end{equation}
we restrict to the region where \( \epsilon(z) > 0 \), since otherwise no carriers contribute.

We evaluate the \( zz \)-component of the conductivity. Using spherical coordinates, we write
\begin{equation}
    d^3\bm{k} = k^2 \sin\theta\, dk\, d\theta\, d\phi, \quad
    v^z = \frac{\hbar k_z}{m^*} = \frac{\hbar k \cos\theta}{m^*}.
\end{equation}
Thus the integrand becomes
\[
\left( v^z \right)^2 = \left( \frac{\hbar k \cos\theta}{m^*} \right)^2 = \frac{\hbar^2 k^2 \cos^2\theta}{(m^*)^2}.
\]

Substituting into the expression, we obtain
\begin{align}
    \sigma^{zz}(z,\omega) &= \frac{2e^2 \tau}{1 - i \omega \tau} \int \frac{d^3\bm{k}}{(2\pi)^3}
    \delta\left( \frac{\hbar^2 k^2}{2 m^*} - \epsilon(z) \right)
    \cdot \frac{\hbar^2 k^2 \cos^2\theta}{(m^*)^2} \notag \\
    &= \frac{2e^2 \tau}{1 - i \omega \tau} \cdot \frac{1}{(2\pi)^3} \cdot \int_0^\infty dk\, k^2\, \delta\left( \frac{\hbar^2 k^2}{2 m^*} - \epsilon(z) \right)
    \cdot \frac{\hbar^2 k^2}{(m^*)^2} \cdot \int_0^\pi d\theta\, \sin\theta \cos^2\theta \cdot \int_0^{2\pi} d\phi.
\end{align}

Evaluating the angular integrals:
\[
\int_0^{2\pi} d\phi = 2\pi, \quad
\int_0^\pi \sin\theta \cos^2\theta\, d\theta = \frac{2}{3},
\]
so the total angular factor is \( \frac{4\pi}{3} \).

Now we handle the radial \( k \)-integral using the delta function:
\[
\delta\left( \frac{\hbar^2 k^2}{2 m^*} - \epsilon(z) \right)
= \frac{m^*}{\hbar^2 k_F(z)} \delta(k - k_F(z)),
\]
so the integral becomes
\[
\int_0^\infty dk\, k^2 \delta(k - k_F) \cdot \frac{\hbar^2 k^2}{(m^*)^2}
\cdot \frac{m^*}{\hbar^2 k} = \frac{k_F^3(z)}{m^*}.
\]

Putting everything together:
\begin{equation}
    \sigma^{zz}(z,\omega) = \frac{2e^2 \tau}{1 - i \omega \tau} \cdot \frac{1}{(2\pi)^3} \cdot \frac{4\pi}{3} \cdot \frac{k_F^3(z)}{m^*}
    = \frac{e^2 \tau}{1 - i\omega\tau} \cdot \frac{k_F^3(z)}{3\pi^2 m^*}
    .
\end{equation}

In terms of the local potential \( U(z) \), this becomes:
\begin{equation}
    \sigma^{zz}(z,\omega) = \frac{e^2 \tau}{1 - i\omega\tau}
    \cdot \frac{1}{3\pi^2 m^*} \left( \frac{2m^*}{\hbar^2} \right)^{3/2} \left( E_F - U(z) \right)^{3/2}
     \quad \text{for } E_F > U(z),
\end{equation}
and \( \sigma^{zz}(z,\omega) = 0 \) otherwise. Then we can impose the $z$-dependent $U$, $m^*$ to get the final results numerically.

\subsection{Numerical Solution of the Steady-State Conductivity Equation}

To compute the steady-state electric potential $\phi(x,z)$ in a graded material with spatially varying local conductivity $\sigma(w)$, where $w = x\sin\theta + z\cos\theta$, we solve the following partial differential equation:
\begin{equation}
\nabla \cdot \left[ \sigma(x,z) \nabla \phi(x,z) \right] = 0,
\end{equation}
subject to Dirichlet boundary conditions
\begin{equation}
\phi(x, z=0) = 1, \qquad \phi(x, z=L) = 0,
\end{equation}
and homogeneous Neumann boundary conditions along the $x$-direction,
\begin{equation}
\left.\frac{\partial \phi}{\partial x}\right|_{x=0} = \left.\frac{\partial \phi}{\partial x}\right|_{x=L_x} = 0.
\end{equation}

\subsubsection{Finite Difference Discretization}

We discretize the domain $[0,L_x] \times [0,L_z]$ using a uniform Cartesian grid with $N_x$ and $N_z$ points in the $x$ and $z$ directions, respectively. The mesh spacings are given by $\Delta x = L_x/(N_x - 1)$ and $\Delta z = L_z/(N_z - 1)$. We denote the discrete potential by $\phi_{i,j} \equiv \phi(x_i, z_j)$.

The conductivity is given by $\sigma(x,z) = \sigma(w)$, with $w = x\sin\theta + z\cos\theta$, and is evaluated at each grid point.

The governing PDE is discretized using a five-point stencil with harmonic averaging of the conductivity at half-grid points. The discrete equation at interior points reads:
\begin{align}
&\frac{1}{\Delta x^2} \left[ \sigma_{i+\frac{1}{2},j} (\phi_{i+1,j} - \phi_{i,j}) - \sigma_{i-\frac{1}{2},j} (\phi_{i,j} - \phi_{i-1,j}) \right] \notag\\
+&\frac{1}{\Delta z^2} \left[ \sigma_{i,j+\frac{1}{2}} (\phi_{i,j+1} - \phi_{i,j}) - \sigma_{i,j-\frac{1}{2}} (\phi_{i,j} - \phi_{i,j-1}) \right] = 0,
\end{align}
where each $\sigma_{i\pm\frac{1}{2},j}$ or $\sigma_{i,j\pm\frac{1}{2}}$ is computed by averaging the values of $\sigma$ at adjacent grid points:
\begin{equation}
\sigma_{i+\frac{1}{2},j} = \frac{1}{2}[\sigma_{i,j} + \sigma_{i+1,j}], \quad
\sigma_{i,j+\frac{1}{2}} = \frac{1}{2}[\sigma_{i,j} + \sigma_{i,j+1}].
\end{equation}

Dirichlet boundary conditions are imposed directly at $z=0$ and $z=L$, while Neumann conditions at $x=0$ and $x=L_x$ are incorporated using one-sided finite differences.

\subsubsection{Linear System and Solution}

The discrete equations are assembled into a sparse linear system $A \vec{\phi} = \vec{b}$, where $\vec{\phi}$ contains the unknown nodal values $\phi_{i,j}$. The resulting system is solved using a direct sparse solver.

\subsubsection{Current Density Evaluation}

Once the potential field $\phi(x,z)$ is obtained, the current density is computed as:
\begin{equation}
J_x = -\sigma(x,z)\frac{\partial \phi}{\partial x}, \qquad
J_z = -\sigma(x,z)\frac{\partial \phi}{\partial z},
\end{equation}
using central difference approximations at interior points. The resulting vector field $\mathbf{J}(x,z)$ represents the local electrical current in the material.

\section{Detailed Numerical Methodology for Graded Diode Simulations}

We solve the nonlinear Poisson equation for one-dimensional graded p-n junctions. This section details the governing equations, discretization and solution method, and the variable-bias and current-density formulation used for I-V curves.

\subsection{Governing equations}

We solve the Poisson equation
\begin{equation}
  \frac{d^2\phi}{dz^2} = -\frac{q}{\epsilon} \left[ p(z) - n(z) + N_d(z) - N_a(z) \right]
\end{equation}

In the non-degenerate limit, carrier densities follow Boltzmann statistics:
\begin{equation}
  n(z) = N_c(z) \exp\left(\frac{E_F + q\phi(z) - U_c(z)}{k_B T}\right)
\end{equation}
\begin{equation}
  p(z) = N_v(z) \exp\left(\frac{U_v(z) - E_F - q\phi(z)}{k_B T}\right)
\end{equation}

with
\begin{align}
  N_c(z) &= \text{prefactor} \times \left[m_c^*(z) k_B T\right]^{3/2} \\
  N_v(z) &= \text{prefactor} \times \left[m_v^*(z) k_B T\right]^{3/2}
\end{align}

The built-in potential is
\begin{equation}
V_{bi} = \frac{k_B T}{q} \ln\left(\frac{N_d(+\infty) N_a(-\infty)}{N_c(+\infty) N_v(-\infty) \exp\left(-\frac{U_c(+\infty) - U_v(-\infty)}{k_B T}\right)}\right)
\end{equation}
with boundary conditions: the left boundary (p-side) is $\varphi(-L/2) = 0$ (reference potential) and the right boundary (n-side) is $\varphi(+L/2) = V_{bi}$ (built-in potential). In the implementation, the computational grid is $z\in[0,L]$ and a shifted coordinate $z_{\rm eff}=z-L/2$ is used; the solution is shifted so that $\varphi(0)=0$ for plotting.

\subsection{Discretization and solution method}

The domain is discretized on a uniform grid with $N = 401$ points and spacing $\Delta z = L/(N-1)$. Using the shifted coordinate $z_{\rm eff}$, we apply a central-difference scheme for the second derivative:
\begin{equation}
\frac{d^2\phi}{dz^2}\bigg|_{z_i} \approx \frac{\phi_{i+1} - 2\phi_i + \phi_{i-1}}{(\Delta z)^2}
\end{equation}
This creates a tridiagonal system for the Laplacian operator.

\subsubsection{Equilibrium (Newton solve)}
At equilibrium ($V_{\rm app}=0$), we solve Poisson's equation coupled to Boltzmann carrier statistics. A charge-neutrality analytic guess is used to initialize $\phi$, and Dirichlet boundary conditions are set from this guess. The nonlinear Poisson equation is solved by Newton--Raphson. At node $i$,
\begin{equation}
R_i(\phi)=\frac{\phi_{i+1}-2\phi_i+\phi_{i-1}}{(\Delta z)^2}
 + \frac{q}{\epsilon}\left[N_{d,i}-N_{a,i}+p_i(\phi_i)-n_i(\phi_i)\right]=0.
\end{equation}
The Jacobian is tridiagonal with diagonal entries
\begin{equation}
J_{ii}=-\frac{2}{(\Delta z)^2}+\frac{q}{\epsilon}\left(\frac{dp_i}{d\phi_i}-\frac{dn_i}{d\phi_i}\right),
\end{equation}
and off-diagonal entries $J_{i,i\pm1}=1/(\Delta z)^2$, where
$dn_i/d\phi_i = n_i(q/k_B T)$ and $dp_i/d\phi_i = -p_i(q/k_B T)$.
We solve $J\Delta\phi=-R$ and update $\phi\leftarrow\phi+\lambda\Delta\phi$ with a backtracking line search
to ensure monotone residual decrease. The equilibrium convergence criterion is $\|R\|_\infty<10^{-8}$.

\subsubsection{Non-equilibrium (Gummel iteration)}
For non-equilibrium, we solve Poisson and the electron/hole continuity equations self-consistently using a Gummel loop.
At each bias $V_{\rm app}$, we iterate:
(i) solve Poisson for $\phi$ with $n,p$ fixed;
(ii) solve the electron continuity equation for $n$ with $\phi,p$ fixed;
(iii) solve the hole continuity equation for $p$ with $\phi,n$ fixed.
Bias-dependent Dirichlet boundary conditions are $\phi(0)=0$ and $\phi(L)=V_{bi}-V_{\rm app}$, and Ohmic contact
carrier densities are enforced using the equilibrium relations at the contacts.

\subsection{Variable bias and current density}

For I-V curve calculation, the applied voltage is swept from $-0.5$ to $+0.5$ V in steps of $0.01$ V. We use a Gummel iteration (Poisson + continuity equations) with the equilibrium solution as initial guess at each bias. The I-V curve is obtained by plotting current versus voltage.

The electron and hole continuity equations in steady state are
\begin{equation}
\frac{dJ_n}{dz}=qR,\qquad -\frac{dJ_p}{dz}=qR,
\end{equation}
with $R$ the recombination rate (SRH in our implementation).

For the continuity solves we use the drift--diffusion form with a Scharfetter--Gummel discretization, which preserves stability for large potential gradients. At each Gummel iteration, the electron and hole continuity equations are solved on the same mesh using the local potentials and band edges. The drift--diffusion currents are
\begin{align}
J_n &= q\mu_n n E_z + q D_n \frac{\partial n}{\partial z},\\
J_p &= q\mu_p p E_z - q D_p \frac{\partial p}{\partial z},
\end{align}
and are discretized using the Bernoulli function $B(x)=x/(e^x-1)$ as
\begin{equation}
J_n = \frac{qD_n}{\Delta z}\left[n_{i+1}B(-\Delta W_n) - n_i B(\Delta W_n)\right],\quad
W_n=\frac{U_c - q\phi}{k_BT}-\ln N_c
\end{equation}
and an analogous expression for $J_p$ with $W_p=\frac{q\phi-U_v}{k_BT}-\ln N_v$. The total current is $J_{\mathrm{total}}=J_n+J_p$.

At each bias point, the Gummel loop is iterated until the maximum change in $\phi$, $n$, and $p$ falls below the tolerance, and the resulting current is evaluated at the device midpoint for the I--V curve.


\end{document}